\RequirePackage{fix-cm}
\documentclass{svjour3}                     
\smartqed  

\usepackage{bookmark}
\usepackage[authoryear]{natbib}
\usepackage{graphicx}
\usepackage{textcomp}
\usepackage{xcolor}
\usepackage{microtype}
\usepackage{booktabs}
\usepackage[mode=text]{siunitx}
\usepackage{etoolbox}
\usepackage{enumitem} 
\usepackage{textgreek}
\usepackage{tipa}
\usepackage{url}
\usepackage{times}
\usepackage{ragged2e}
\usepackage{caption}
\usepackage{times}

\newcolumntype{P}[1]{>{\raggedright\arraybackslash}p{#1}}
\newcolumntype{R}[1]{>{\raggedleft\arraybackslash}p{#1}}
\usepackage{makecell}

\usepackage[
paperwidth=15.51cm,
paperheight=23.52cm,
margin=0.65in
]{geometry}

\usepackage{hyperref}
\hypersetup{
 colorlinks=true,   
 citecolor=blue,
 urlcolor=blue,
 linkcolor=blue,
 bookmarksnumbered=true,     
 bookmarksopen=true,         
 bookmarksopenlevel=1,       
 colorlinks=true,            
 pdfstartview=Fit,           
 pdfpagemode=UseOutlines,    
 pdfpagelayout=TwoPageRight 
}

\usepackage{titlesec}

\usepackage[lining]{FiraSans} 
\usepackage[T1]{fontenc}

\titleformat{\section}
  {\sffamily\bfseries\fontsize{11}{13}\selectfont}{\thesection}{1em}{}

\titleformat{\subsection}
  {\sffamily\bfseries\fontsize{10}{12}\selectfont}{\thesubsection}{1em}{}

\titleformat{\subsubsection}
  {\sffamily\bfseries\itshape\fontsize{9}{11}\selectfont}{\thesubsubsection}{0.5em}{}

\begin{document}

\title{From Anecdote to Evidence: The Relationship Between Personality and Need for Cognition of Developers}
\titlerunning{The Relationship Between Personality and Need for Cognition of Developers}

\author{Daniel Russo \and Andres R. Masegosa \and Klaas-Jan Stol}


\institute{D. Russo \at
Department of Computer Science, Aalborg University, Copenhagen, Denmark\\
              \email{daniel.russo@cs.aau.dk}           
\and
A. R. Masegosa \at
Department of Computer Science, Aalborg University, Copenhagen, Denmark\\
              \email{arma@cs.aau.dk}           
           \and
            K.-J. Stol \at
            Lero---The Science Foundation Ireland Research Centre for Software and\\
School of Computer Science \& Information Technology,
University College Cork, Cork, Ireland \\
            \email{k.stol@ucc.ie}
}

\date{Received: date / Accepted: date}

\maketitle

\begin{abstract}
There is considerable anecdotal evidence suggesting that software engineers enjoy engaging in solving puzzles and other cognitive efforts. A tendency to engage in and enjoy effortful thinking is referred to as a person's `need for cognition.'
In this article we study the relationship between software engineers' personality traits and their need for cognition. Through a large-scale sample study of 483 respondents we collected data to capture the six `bright' personality traits of the HEXACO model of personality, and three `dark' personality traits. 
Data were analyzed using several methods including a multiple Bayesian linear regression analysis.
The results indicate that ca. 33\% of variation in developers' need for cognition can be explained by personality traits. 
The Bayesian analysis suggests four traits to be of particular interest in predicting need for cognition: openness to experience, conscientiousness, honesty-humility, and emotionality. Further, we also find that need for cognition of software engineers is, on average, higher than in the general population, based on a comparison with prior studies.
Given the importance of human factors for software engineers' performance in general, and problem solving skills in particular, our findings suggest several implications for recruitment, working behavior, and teaming.
\end{abstract}

\section{Introduction}


\begin{quote}
    ``You have to simulate in your mind how the program's going to work, and you have to have a complete grasp of how the various pieces of the program work together.'' ---Bill Gates \cite[p. 73]{lammers1986} 
\end{quote}

There is a considerable body of research that has studied software engineers as ``problem solvers,'' that seeks to understand how developers make sense of problems and how they solve design puzzles \citep{petre1997}.
The sense of accomplishment from solving problems and programming activities is often seen as a critical characteristic of developers \citep{almstrum2003attraction}.
Similarly, a review article on motivation factors of software engineers concludes that: 
\begin{quote}
``learning, exploring new techniques and problem solving appear to be the motivating aspects of SE. However little work has focused on the specific nature of Software Engineering itself [...] neither did any of the models focus on the nature of the SE's job itself such as [...] the logical nature of problem solving, use of creativity, complex problem solving, and so on.'' 
\cite[p. 874]{beecham2008motivation}
\end{quote}

Beecham et al. suggest that a software engineer's job involves problem solving and use of creativity---in other words, cognitive tasks.
The role of cognition and, in particular, the ability to solve problems in programming has been acknowledged as an important factor in developer productivity since the early days of computing \citep{curtis1984fifteen}.
One key psychological attribute for software engineers is the joy of effortful cognitive activities, or what has been termed \textit{need for cognition} (NfC) \citep{cacioppo1982need}. The level of need for cognition indicates the extent to which a developer enjoys cognitive effort, which is directly related to problem solving skills \citep{heppner1983cognitive}.
Understanding what drives developers' need for cognition can be 
useful to software development organizations, for example, in hiring decisions and onboarding new staff \citep{sharma2020}. 
Despite decades of research and an increasing level of attention for human aspects in software engineering \citep{cruz2015forty,curtis1984fifteen,lenberg2015behavioral,storey2020}, need for cognition of software engineers has, as of yet, not been studied (cf. \cite{capretz2003}). 
Given the importance of the role that need for cognition plays in software engineers' problem solving skills, it is surprising that there is a paucity of research that has focused on this construct. 
Prior studies that focused on software development tasks and developer preferences have sought explanations in developers' personality traits.
For example, studies have linked  personality traits to developers' ability in code review \citep{da2007does}, exploratory testing \citep{shoaib2009empirical}, object-oriented programming \citep{cegielski2006makes}, and programming proficiency \citep{evans1989best}. 
Similarly, other studies highlight the relation between developer behavior and attitudes,  and personality traits. For example, decision-making \citep{feldt2010links}, working attitude \citep{smith1989personality}, or preferences for using offline batch versus online time-sharing equipment \citep{lee1978personality}. More recently, \cite{mellblom2019} found a strong link between neuroticism (one of the personality traits in the Five Factor Model that is also used in this study) and developer burnout.
These studies suggest that personality traits theory can be an insightful theoretical perspective to study software developers. 
Hence, in this study we adopt this perspective as we seek to find empirical evidence, rather than relying on anecdote, to establish developers' need for cognition. Specifically, we address the following research question:

\begin{quote}
    \textbf{Research Question}: What is the relationship between personality traits of software engineers and their need for cognition?
\end{quote}

To answer this question, we conducted a large-scale sample study of almost 500 software professionals. We analyzed this rich data set using Bayesian analysis procedures.
Based on the findings, we offer recommendations for software organizations for attracting and retaining software professionals.

This article is structured as follows. 
Section~\ref{sec:related} provides an overview of prior studies of personality traits of software developers.
Sec.~\ref{sec:design} presents the research design.
Afterward, in Sec.~\ref{sec:analysis}, we present the results of our analysis. 
Findings with recommendations are discussed in Sec.~\ref{sec:discussion}. Finally, Sec.~\ref{sec:conclusion} concludes the paper and offers suggestions for future research.

\section{Background and Related Work}
\label{sec:related}

Software engineering is a socio-technical field of research \citep{perry1994people}, which seeks to advance the state of practice of software development. 
While historically the software engineering field has had a strong focus on the technical aspects, there has been a growing interest in studying human factors in software engineering \citep{boehm1988understanding,glass2002facts,lenberg2015behavioral,feldt2010links,storey2020}. 


To explain developer preferences and behavior, several studies have considered personality traits \citep{weinberg1971psychology,lee1978personality,shneiderman1980software}.
For example, \cite{li2020distinguishes} provides ``18 attributes pertained to engineers’ personalities'' by asking developers to rank 54 attributes.
\cite{cruz2015forty} identified around 90 articles on personality research in software engineering, published between 1970 and 2010.
However, the interest for personality studies has surged dramatically in the last 20 years or so.  Over 70\% of studies reviewed by \cite{cruz2015forty} were published after 2002.

Need for cognition \citep{cacioppo1982need} is a critical psychological attribute for software professionals, and  related to problem-solving skills \citep{heppner1983cognitive},
which are essential for productive software engineers \citep{graziotin2014happy}.
We identified only three studies that discussed the need for cognition related to computer applications, though none of them studied software professionals specifically.
The most recent study concludes through a small-scale laboratory experiment involving 18 undergraduate students that using a web development learning platform improves students' need for cognition \citep{bergande2020codetripping}.
However, since several previous studies in social psychology  through investigations involving a large number of subjects suggest that need for cognition is a stable trait \citep{furnham2013stable,roberts2008stable,bruinsma2018longitudinal}, we suggest treating the findings of that laboratory experiment with care.
Another study by \cite{amichai2007effects} tested several hypotheses related to the level of need for cognition and internet use.
Findings from a laboratory experiment with 182 participants suggest that people use the same number of hyperlinks regardless of their level of need for cognition; that individuals scoring high in NfC are less influenced by interactive or aesthetic components; and that the time spent on the internet is the same for both groups.
Finally, \cite{van2006paradox} conducted a laboratory experiment involving 43 participants and demonstrated that the usability of Graphical User Interfaces does not depend on the users' NfC.
We identified only one study that linked NfC to personality traits \citep{sadowski1997need}, but like the studies discussed above, this study also relied on a sample of participants from the general population rather than software engineers.
In this study, we seek to investigate the relationship between personality traits and need for cognition that specifically focuses on software developers. 
Studying software developers  will allow us to draw comparisons with studies using samples of a general population, and to identify specific implications for a software engineering context.

\subsection{Personality Traits}

Personality can be considered as a set of patterns of thinking, feeling, and behaving, which are based on a set of traits that are predictors of a person's behavior and action \citep{corr2009personality}.
Karl Gustav Jung first introduced personality traits-based theories in the early 1930s \citep{jung1931basic}.
These theories were later operationalized employing measurement instruments, such as the Meyers-Briggs type indicators (MBTI) \citep{myers1976introduction}.
At least until 2010, the MBTI was the most widely used personality test within software engineering research \citep{cruz2015forty}. 

However, the Myers-Briggs classification has been criticized for its lack of validity and utility, such as unstable test-retest reliability and inaccurate predictive validity \citep{boyle1995myers}. 
Moreover, the MBTI is not correlated with other personality scales \citep{furnham1996big}, and has not been validated by empirical research \citep{druckman1991mind}. 
Psychology researchers observed that the Meyers-Briggs type indicators do not measure categorically distinct types, generating quasi-random traits assignments~\citep{mccrae1989reinterpreting}.
The reason is that the MBTI is based on a dichotomous preference score and not a continuous one, which limits the extent to which personality traits can be measured~\citep{devito1985review}.

To address these limitations, McCrae and colleagues developed the Five-Factor Model (FFM), which is grounded in a series of empirical evaluations \citep{mccrae1989reinterpreting,mccrae1992introduction}.
The FFM is also known as the OCEAN model, an acronym for the five  personality traits it measures: Openness (O), Conscientiousness (C), Extraversion (E), Agreeableness (A), and Neuroticism\footnote{The term `Neuroticism' should not be used as such since it might suggest a mental health issue. Emotionality or Emotional stability are better terms.} (N).
\cite{ashton2004six} have extended the OCEAN model by including an additional trait: Honesty-Humility. 
This trait emerged from large-scale studies that identified it as a distinct personality trait \citep{saucier2009recurrent}.
The HEXACO model has become the new standard in social psychology \citep{ashton2014hexaco}. 
Replication studies of the OCEAN model (which used more advanced computation techniques that were not available when the OCEAN model was initially proposed) confirmed the validity of this new trait \citep{ashton2013individual}.

The characteristics (or \textit{facets}) of the six personality traits of the HEXACO model are: 

\begin{enumerate}
\item Honesty-Humility (H-H): sincerity, fairness, greed avoidance, and modesty.
\item Emotionality (Emo): fearfulness, anxiety, dependence, and sentimentality.
\item Extraversion (Ext): social self-esteem, social boldness, sociability, and liveliness.
\item Agreeableness (Agr): forgivingness, gentleness, flexibility, and patience.
\item Conscientiousness (Con): organization, diligence, perfectionism, and prudence.
\item Openness to Experience (OtE): aesthetic appreciation, inquisitiveness, creativity, and unconventionality.
\end{enumerate}

Personality traits can be classified as either `bright' or `dark' \citep{judge2009bright}.
The HEXACO (as also the OCEAN) model focuses on bright traits, i.e. those traits that are considered ``desirable.''
High scores (low scores only for Emotional stability, since it measures the tendency to experience danger) indicate bright personality traits.  

There are also dark personality traits; that is, high scores on these traits suggest less desirable characteristics. 
Commonly used in personality studies is the so-called ``Dark Triad'' comprising
Narcissism, Psychopathy, and Machiavellianism, which are defined as follows \citep{paulhus2002dark}:

\begin{enumerate}
\item Narcissism (Nar): is related to the individual gratification through self-centered, selfish, and egoistic behaviors.
\item Psychopathy (Psy): anti-social behaviors such as lack of empathy or remorse characterize this trait. 
\item Machiavellianism (Mac): is named after Niccolo Machiavelli (author of \textit{The Prince} \citep{machiavelli1532}), focusing on the manipulation to gain self-interested goals and readiness to exploit other people to achieve them. 
\end{enumerate}

\subsection{Need for Cognition}

Need for cognition is a psychological attribute advanced in the early 1980s by \cite{cacioppo1982need}, and refers to an individual's tendency to engage and enjoy effortful thinking.
This construct was developed to assess differences among individuals of cognitive processing \citep{sadowski1997need}.
Individuals with high scores on this construct have positive attitudes toward problem-solving, reasoning, and abstraction \citep{cacioppo1996dispositional}.
People scoring low rely on cognitive heuristics and stereotypes to interpret reality \citep{leary2009handbook}.

Individual psychological attributes are measurable proxies of behavioral characteristics \citep{salzberger2013attempting}, such as intelligence and satisfaction with life \citep{kosinski2013private}.
Personality traits are also defined as a set of psychological attributes, which are able to predict other attributes, for example cognitive ability  \citep{roberts2007power}.
In this study we assess the link between personality traits and need for cognition.

It is important to emphasize that need for cognition does not represent a person's intelligence, although some studies found a positive correlation between these two constructs \citep{hill2013need}. A key difference is that need for cognition  only reflects information processing motivations, and not cognitive abilities \citep{cacioppo1996dispositional}. 
Notwithstanding, need for cognition is related to other attributes, such as cognitive and sensory innovativeness, which refers to the joy to engage in new experiences to ``stimulate the mind'' \citep{venkatraman1990differentiating}, attributional complexity (i.e., the level of interest or motivation about reality, a preference for complex over simple explanations, meta-cognition capabilities, abstraction skills, and a tendency to infer abstract or complex attributions) \citep{fletcher1986attributional}, and a high level of tolerance for ambiguity and cognitive flexibility \citep{petty1996individual}.

\cite{tuten2001understanding} noted that need for cognition is usually correlated with certain personality traits, in particular with openness to experience (defined above).
However, scholars have pointed out that need for cognition is a theoretically distinct construct: an individual might be open to new experiences \textit{without} reflecting on the consequences of such new experiences \citep{kearney2009and}.
On the other hand, introverts who are reluctant to new experiences might enjoy thinking in a reassuring environment, such as reading an interesting book in the comfort of their home.
In fact, extraversion (the tendency to enjoy human interactions) is not related to need for cognition \citep{sadowski1997need}.
Therefore, regardless of the fact whether personality traits are correlated or not with need of cognition, all those personality attributes have unique meanings and are thus, considered separately \citep{mccrae1996social}. 

Need for cognition is also a relevant predictor of positive behaviors in working environments.
Research suggests that a high level of need for cognition is a robust negative moderator of social loafing (the tendency to exert less effort when working in groups), showing significantly higher team efforts \citep{smith2001individual}.
Further, people with high need for cognition values tend to exhibit higher job satisfaction if they face complex engineering tasks, and are generally more satisfied with their jobs than employees scoring low \citep{park2008need}.
Finally, the need for cognition reduces depressed moods at work \citep{gallagher2012managing}. 

Anecdotal evidence suggests that software engineers would score high on need for cognition, but there are no studies that have empirically investigated this. Given the socio-technical nature of software development and the importance of human aspects in software engineering, we set out a large-scale study to investigate need for cognition among software engineers.

\section{Research Design}
\label{sec:design}
To investigate the need for cognition attribute of software developers, we conducted a large-scale sample study. 
Sample survey research is highly suitable to seek answers that can be generalized to a larger population \citep{stol2018abc}.
In this study we seek to understand the relationship between personality traits and need for cognition of software engineers, considering both bright and dark personality traits.

\subsection{Measurement Instrument}

All constructs in the study were defined as latent variables which we measured using existing instruments that we adopted from the psychology literature. 
To measure the bright personality traits, we adopted the HEXACO model of personality \citep{ashton2004six}.
While different survey instruments exist for measuring traits, we adopted the Brief HEXACO Inventory (BHI), which consists of 24 questions \citep{de2013BHI}.
This instrument has shown a high level of stability and accuracy; moreover, it correlates well with other similar and longer instruments (e.g., the HEXACO-PI-R) \citep{de2013BHI}.
To measure the dark traits, we adopted the so-called ``Dirty Dozen'' \citep{jonason2010dirty}, which is also a widely used survey instrument in psychology research \citep{rauthmann2012dark}.
Need for cognition was measured using the NCS-6 instrument, which is a parsimonious instrument with six items \citep{lins2018very}.
We followed the instrument guidelines (which indicate which items are reverse-coded). 
We calculated factor scores (which represent a respondent's score on a particular trait) using the mean function; which is a standard procedure when using frequently used instruments such as this \citep{weiner2017handbook}.

\subsection{Data Collection and Screening}

To collect data we used the Prolific data collection platform, which is a dedicated platform for academic research (www.prolific.co) with over 88,000 active members worldwide at the time of the survey. 
The use of Prolific for data collection has several advantages over other data collection procedures (for example, invitations through mailing lists), such as reliability, replicability, and data quality \citep{peer2017beyond,palan2018prolific}.
Prolific is widely used in other research fields such as economics \citep{marreiros2017now}, psychology \citep{callan2017interrelations}, and food science \citep{simmonds2018show}.
More recently, studies within computer science and software engineering have adopted the Prolific as a data collection platform \citep{Hosio2020CrowdsourcingDiets,van2021effect,russo2020predictors,russo2021agile_success_model,russo2021daily,cucolas2021impact,russo2021understanding}.
To implement the survey questionnaire, we used Qualtrics (www.qualtrics.com). 
To minimize response bias, we randomized questions within their blocks \citep{paulhus1991,gravetter2018research}.

The choice of sampling strategy is key for sample research.
\cite{baltes2020sampling} recently suggested that most studies in software engineering research use non-probability sampling, such as purposive or convenience sampling, and less than 10\% use random sampling. 
We selected a cluster-randomized sampling strategy \citep{gravetter2018research}: a number of potential respondents were selected at random from the Prolific population.
In other words, we did not selected the entire world population, but just a cluster of it (i.e., the Prolific community).
From our cluster we proceeded to identify software professionals through a multi-stage selection process described below.
Compared to random sampling, this is less precise (since \textit{only} a portion of the population is considered, in this case, that portion which has a presence on Prolific), but it is very cost-effective compared to the other strategies.
To be sure to include only software developers in our sample, we ran a three-step screening process as detailed below.

\paragraph{Step 1. Pre-Screening.} The first step was a pre-screening of the active members on Prolific based on some general criteria related to software development, such as knowledge of software development techniques, doing computer programming for a living, and the use technology at work. Further, we required a Prolific platform approval rate of 100\%, suggesting that all included respondents successfully completed prior surveys (and did not fail any screening criteria in previous surveys).
After pre-screening, our sample included 2,897 people.

\paragraph{Step 2. Competence Screening.} In the second step we invited candidates who self-identified as software engineers to take part in a competence screening to ensure that they were sufficiently knowledgeable in the subject matter. 
We asked three multiple-choice questions,\footnote{Questions are included in the supplementary material.} one about software design and two about programming, and also time-boxed the responses within three minutes to avoid suspicious behavior and cheating. 
At the end of this stage, we had a sample of 514 participants.

\paragraph{Step 3. Quality Screening.} At this point we administered the survey to the remaining participants, but included a quality screening to ensure high-quality data. The quality screening comprised three attention checks; participants who did not pass these checks were excluded from the sample.
As a final result of the entire screening process, we received 483 valid and complete responses that could be used for analysis.

Ethical issues were considered according to the recommendations in the Declaration of Helsinki \citep{general2014Helsinki}.
Participants were identified only through a user ID provided by the Prolific data collection platform.
We did not collect any potentially sensitive information of participants. 
Further, participants were free to withdraw their participation at any time up to the point of submitting, and some (\textit{n}=306) decided not to participate to this investigation although invited. 
Finally, several of the authors have had research ethics and integrity training, which was specifically tailored to behavioral sciences.

\subsection{Sample Description}\label{sub:sample}
\begin{figure}[!ht]

\includegraphics[width=2.5in,trim={0 2cm 0 0cm},clip]{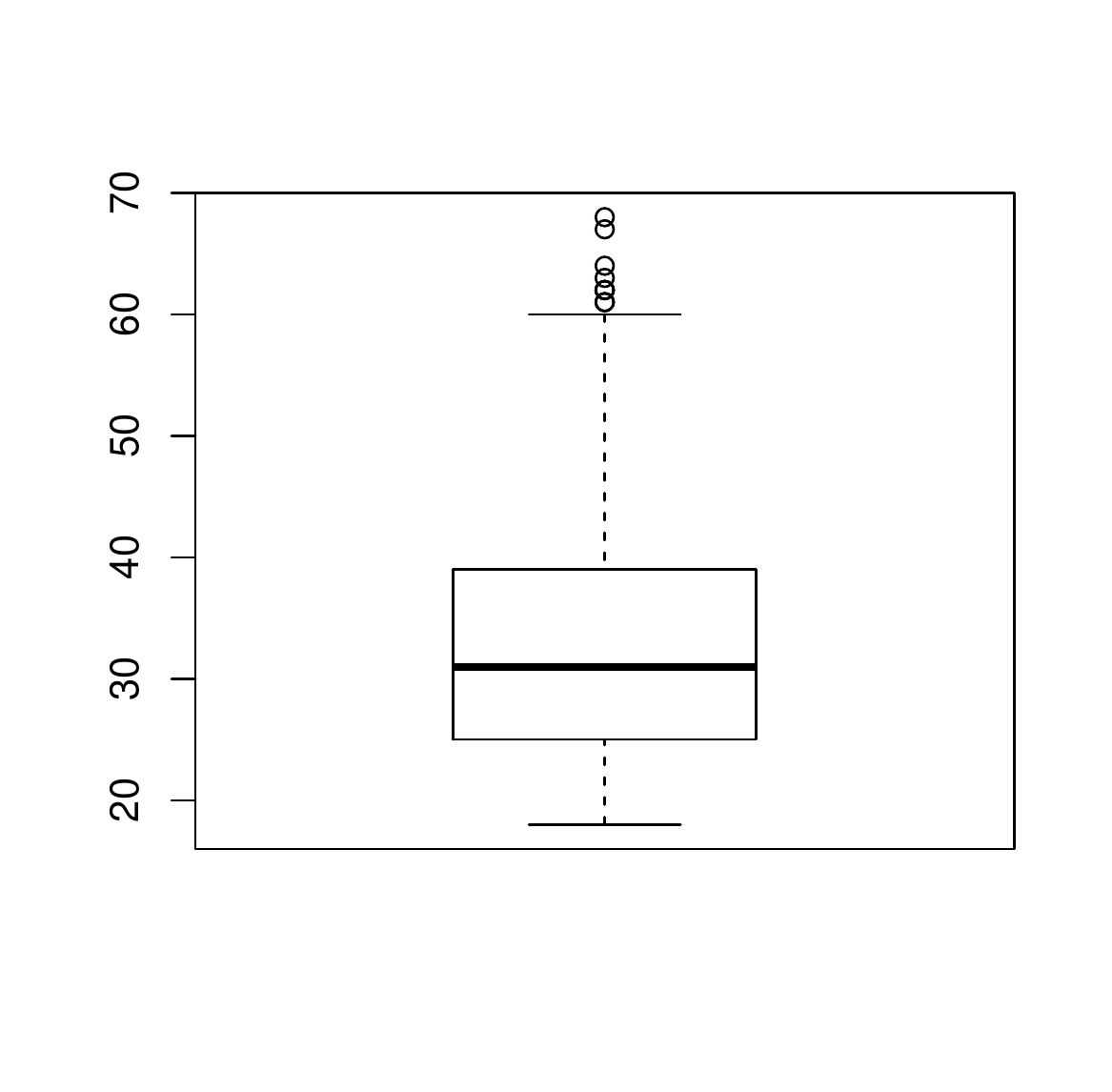}
\caption{Distribution of participants' age}
\label{fig:age}
\end{figure}

The final sample included 483 complete responses, representing the largest sample for a personality research study in software engineering \citep{cruz2015forty}.
In addition to data on personality traits, we also collected demographic data.
Respondents were between 18 and 68 years of age, with an average of 33 years (see Fig.~\ref{fig:age}).
Table~\ref{tab:gender} shows that almost 19\% of  respondents were women.
Most of the respondents (ca. 93\%) were born in Western countries.
Table~\ref{tab:Country} lists the 15 most frequent countries of origin (representing approximately 90\% of all subjects).
Almost three-quarters of the selected professionals have at least one university degree (Table~\ref{tab:Education}).

\begin{table}[!t]
\sisetup{
    group-digits=true,
    group-minimum-digits=4,
    mode=text,
    detect-weight=true, 
    detect-family=true
}
\caption{Gender distribution of the sample}
\label{tab:gender}
\begin{tabular}{m{3cm}SS}
\toprule
Gender & {Frequency} & {Percent} \\ \midrule
Men & 393 & 81.4 \\ 
Women & 90 & 18.6 \\ 
Total & 483 & 100 \\ \bottomrule
\end{tabular}
\end{table}

\begin{table}[!ht]
\sisetup{
    group-digits=true,
    group-minimum-digits=4,
    mode=text,
    detect-weight=true, 
    detect-family=true
}
\caption{Respondents' country of origin}
\label{tab:Country}
\begin{tabular}{m{4cm}SS}
\toprule
Country & {Frequency} & {Percent} \\ \midrule
United Kingdom & 141 & 29.2 \\ 
USA & 135 & 28.0 \\ 
Portugal & 33 & 6.8 \\ 
Poland & 22 & 4.6 \\ 
Italy & 18 & 3.7 \\ 
Canada & 15 & 3.1 \\ 
Germany & 12 & 2.5 \\ 
Spain & 9 & 1.9 \\ 
Ireland & 9 & 1.9 \\ 
Greece & 8 & 1.7 \\ 
Mexico & 8 & 1.7 \\ 
Australia & 7 & 1.4 \\ 
France & 6 & 1.2 \\ 
Hungary & 5 & 1.0 \\ 
Estonia & 4 & 0.8 \\ 
Other & 51 & 10.5 \\ \bottomrule
\end{tabular}
\end{table}

\begin{table}[!ht]
\sisetup{
    group-digits=true,
    group-minimum-digits=4,
    mode=text,
    detect-weight=true, 
    detect-family=true
}
\caption{Highest degree of education of respondents}
\label{tab:Education}
\begin{tabular}{m{4cm}SS}
\toprule
Education & {Frequency} & {Percent} \\ \midrule
Bachelor's degree & 241 & 49.9 \\ 
Master's degree & 105 & 21.7 \\ 
Some college but no degree & 77 & 15.9 \\ 
High school graduate & 37 & 7.7 \\ 
Doctoral degree & 17 & 3.5 \\ 
Less than high school degree & 2 & 0.4 \\ 
Other & 4 & 0.8 \\ \bottomrule
\end{tabular}
\end{table}

\begin{table}[!ht]

\sisetup{
    group-digits=true,
    group-minimum-digits=4,
    mode=text,
    detect-weight=true, 
    detect-family=true
}
\caption{Respondents' roles within their organizations}
\label{tab:Role}
\begin{tabular}{m{5.4cm}SS}
\toprule
Role &  {Freq.} & {Percent} \\ \midrule
Software developer, programmer & 252 & 52.2 \\ 
Data analyst, engineer, scientist & 44 & 9.1 \\ 
Technical support & 32 & 6.6 \\ 
Team Lead & 31 & 6.4 \\ 
DevOps engineer, infrastructure developer & 22 & 4.6 \\ 
Product manager & 21 & 4.3 \\ 
Tester, QA engineer & 16 & 3.3 \\ 
Architect & 12 & 2.5 \\ 
CIO, CEO, CTO & 12 & 2.5 \\ 
Systems analyst & 12 & 2.5 \\ 
UX, UI designer & 9 & 1.9 \\ 
Other & 20 & 4.1 \\ \bottomrule
\end{tabular}
\end{table}


\section{Bayesian Data Analysis}
\label{sec:analysis}

We conducted the analyses using several software packages. After collecting the data, we performed several sanity checks using IBM SPSS (version 26).
The primary analyses were conducted using the JASP (version 0.11.1) statistical package \citep{jasp2018jasp}. JASP is a relatively new open source package that is actively being developed, and seeks to offer a user-friendly interface to conduct Bayesian analyses. 
We modeled personality traits as exogenous variables (or predictors) and need for cognition as the endogenous (or criterion) variable.

The interest in Bayesian statistics is rising in the empirical software engineering research community \citep{furia2017,furia2019bayesian,furia2021}.
One reason for this is an increasing awareness of the limitations of `traditional' frequentist statistics, which relies on \textit{p}-values and significance testing of null-hypotheses \citep{cohen1994earth,baker2016statisticians}.
By convention, \textit{p}-values are considered as a measure of evidence in favor (or against) of a null hypothesis. Small \textit{p}-values indicate strong evidence against the null hypothesis, and \textit{p}-values lower than 0.05 are widely considered as the critical threshold that determines when we have sufficiently strong evidence against the null hypothesis. Nevertheless, \textit{p}-values cannot be interpreted as the probability that the null hypothesis is true. What they provide is the probability of having data at least as extreme as the one observed, if the null hypothesis was indeed true.\footnote{Also, the threshold lower than 0.05 for the \textit{p}-values is a convention that has to be corrected based on the number of predictors to avoid Type I errors \citep{ACMStd,russo2020predictors}.} Consequently, the use of \textit{p}-values as a measure of strength of the evidence to support any claims must be done with caution \citep{nuzzo2014scientific}. In recent years, scholars of statistics have stressed these arguments and advocated for the use of Bayesian approaches instead of frequentist ones, since the former can offer a more comprehensible, nuanced understanding of the statistical data analysis \citep{wagenmakers2018bayesian,dienes2014Bayes,lee2018determining,furia2019bayesian,JASP2019guidelines}.

Bayesian statistics are not based on a dichotomous evaluation mode (i.e., hypotheses are not accepted or rejected). Instead, the Bayesian  paradigm accepts that different explanations may be possible and that evidence favoring one or another hypothesis can be properly quantified. Bayesian methods allow the computing of the probability that a given hypothesis is true.\footnote{When the set of considered hypothesis does not contain a hypothesis that perfectly represents the underlying data generation process, this probability should be interpreted as the probability that a given hypothesis is the best among all considered hypotheses \citep{masegosa2020learning}.} This probability is the so-called \textit{Bayesian posterior probability}, which is denoted as P(H\textpipe{}data). This probability  quantifies the strength of our evidence in favor of \textit{H}. Bayesian posterior probabilities are computed by applying the Bayes rule over the following two quantities: the observed data's prior probability and the likelihood. Prior probabilities allow us to integrate any prior evidence that we have in favor of a given hypothesis. However, non-informative (usually uniform) probabilities can also be used when little or no prior information is available. The likelihood of the data can be easily understood in terms of Bayes factors. A Bayes factor, defined for two competing hypotheses and noted as BF\textsubscript{10}, represents the observed evidence in favor of \textit{H}\textsubscript{1} over \textit{H}\textsubscript{0}.\footnote{An alternative notation is BF\textsubscript{01}, representing the evidence in favor of \textit{H}\textsubscript{0} over \textit{H}\textsubscript{1}.} Table~\ref{tab:bayesian} presents heuristics to interpret Bayes factors. Rather than delineating explicit thresholds or cut-off values, beyond which a conclusion is supported, as is typical in frequentist statistics, Bayes factors should be interpreted more qualitatively. For example, an analysis resulting in a Bayes factor BF\textsubscript{10} of, say, 5, suggests moderate evidence for \textit{H}\textsubscript{1} over \textit{H}\textsubscript{0}. If the Bayes factor is 50, on the other hand, the heuristics suggest that there is robust evidence for \textit{H}\textsubscript{1} over \textit{H}\textsubscript{0}. Consequently, the `thresholds' in Table~\ref{tab:bayesian} should not be interpreted strictly but rather  heuristically, as a continuum representing strength of evidence. Finally, the following equation described how Bayesian posterior probabilities, prior probabilities, and Bayes factors relate:

\[\frac{P(H_1|D)}{P(H_0|D)} = \frac{P(H_1)}{P(H_0)} \cdot BF_{10} \]

\noindent For example, if \textit{H}\textsubscript{1} and \textit{H}\textsubscript{0} are, a priori, equally probable (i.e., $\frac{P(H_1)}{P(H_0)}=1$), but the observed data suggests strong evidence for \textit{H}\textsubscript{1} with a BF\textsubscript{10} equal to 20, then we have that $\frac{P(H_1|D)}{P(H_0|D)} = 20$. And, solving the equation, we have that \textit{P(H\textsubscript{1}|D)=$\frac{20}{1+20}\approx0.95$} and \textit{P(H\textsubscript{0}|D)=$\frac{1}{1+20}\approx0.05$}.

One of the critical advantages of Bayesian statistics is that the above reasoning can be naturally extended to multiple hypotheses or to more complex settings like multi-model linear regression (as we do in the next section). Because, in Bayesian statistics, all models and observations are equally treated as random variables. So, all computations are always done following standard probability operations, like the ones described above.

\begin{table}[!t]
\sisetup{
    group-digits=true,
    group-minimum-digits=4,
    table-format=0.3,
    mode=text,
    detect-weight=true, 
    detect-family=true
}
\caption{Heuristics for interpretation of Bayes factors BF$_{10}$ \cite[p. 105]{lee2013}}
\label{tab:bayesian}
\begin{tabular}{p{3cm}p{5cm}}
\toprule
Bayes factor & Evidence category \\ 
\midrule
$>$ 100 & Extreme evidence for H$_1$\\
30 -- 100& Very strong evidence for H$_1$\\
10 -- 30& Strong evidence for H$_1$\\
3 -- 10& Moderate evidence for H$_1$\\
1 -- 3& Anecdotal evidence for H$_1$\\
1& No evidence\\
1/3 -- 1& Anecdotal evidence for H$_0$\\
1/10 -- 1/3& Moderate evidence for H$_0$\\
1/30 -- 1/10& Strong evidence for H$_0$\\
1/100 -- 1/30  & Very strong evidence for H$_0$\\
$<$ 1/100& Extreme evidence for H$_0$\\
\bottomrule
\end{tabular}
\end{table}

\subsection{Bayesian multi-model linear regression}\label{sec:BMMLR}

In this study we use a Bayesian multi-model linear regression. Under this model, we assume that the nine personality traits studied here (a.k.a. the independent variables) have a linear relationship with need for cognition (a.k.a. the dependent variable). But, crucially, this analysis assumes that not necessarily all the personality traits are needed to predict developers' need for cognition. In consequence, we explore all regression models defined by all possible combinations of the different personality traits. As we are considering nine traits, this analysis explores 2\textsuperscript{9}=512 alternative regressions models. Each model is treated as a different hypothesis.

We followed the standard guidelines reported by \cite{JASP2019guidelines} and \cite{van2021tutorial} for  conducting the Bayesian data analysis. We first performed several checks to verify that the use of a linear regression model to study the relationship between the considered personality traits and need for cognition is a reasonable choice. Following \cite{van2021tutorial}, we assessed the assumptions of normality, linearity, homoscedasticity, and the absence of multicollinearity. We discuss the procedures for assessing these next. 

\begin{figure}[!b]
\includegraphics[trim={4cm 0cm 0cm 0cm},clip,width=3.5in]{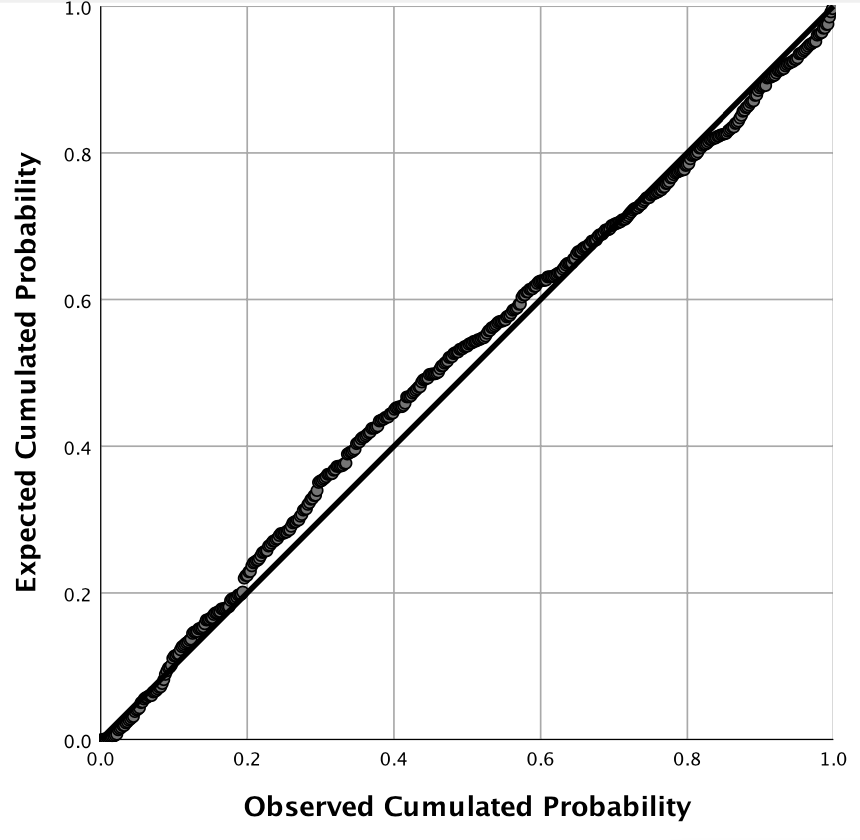}
\caption{Normal P-P plot of regression standardized residuals for need for cognition.}
\label{fig:PPplot}
\end{figure}

\textbf{Data Normality.} We first evaluated data normality. 
Residuals can be seen as the differences between the observed value of need for cognition (i.e., the endogenous variable) and the personality trait values (i.e., the exogenous variables).
Figure~\ref{fig:PPplot} presents a normal predicted probability plot (P-P plot), which shows that the residuals are normally distributed since they are reasonably aligned to the diagonal normality line.

\begin{figure}[!t]
\includegraphics[width=4in]{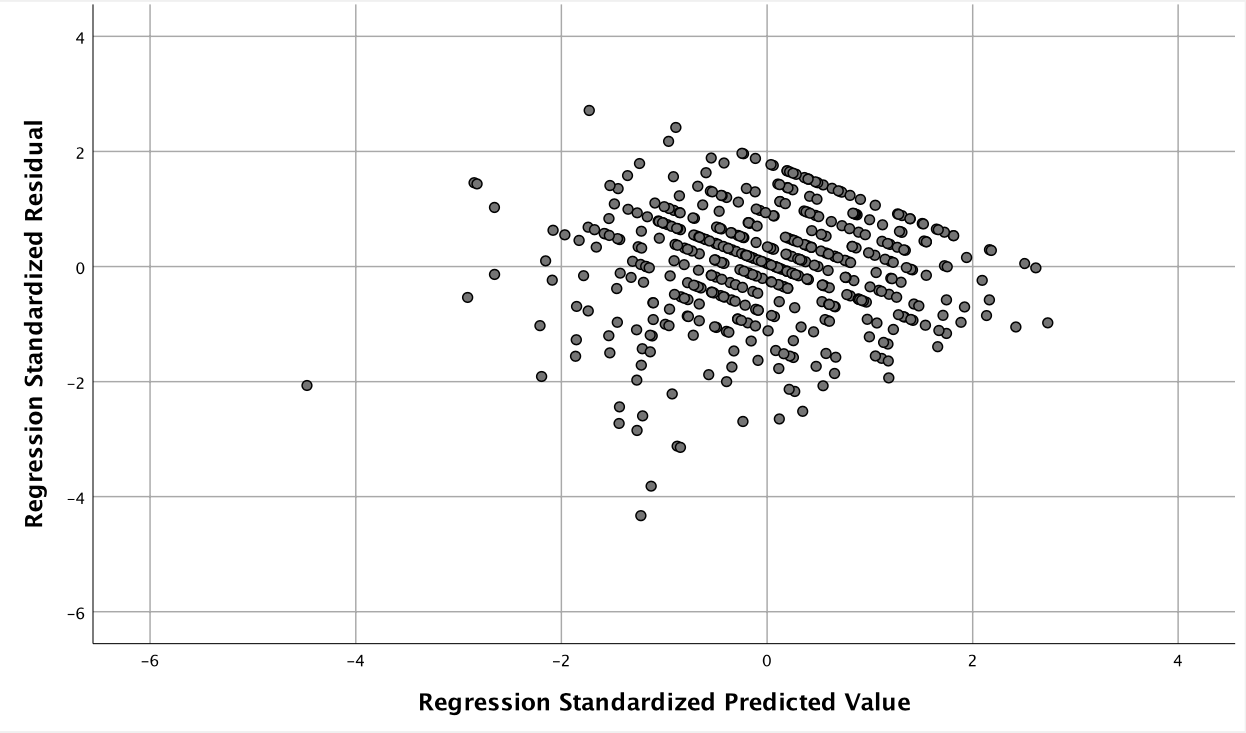}
\caption{Scatterplot of the residuals to assess homoscedasticity for need for cognition}
\label{fig:ResidualsScatterplot}
\end{figure}

\textbf{Homoscedasticity.} Second, we evaluated homoscedasticity by checking whether the residuals are equally distributed.
Heteroscedasticity means that residuals are unevenly spread, leading to clustering on some values. This is undesirable as it would mean that the variance of the observations is not equal.  
The scatterplot shown in Figure~\ref{fig:ResidualsScatterplot} shows that the residuals are randomly distributed. 

Since our residuals are normally distributed and homoscedastic, we can also assume that the exogenous variables (i.e., personality traits) in the regression have a straight-line relationship with the endogenous variable (i.e., need for cognition), concluding that the data have a linear relationship. 
It is linear because the residuals appear to be linear since we do not see any curved pattern in Figure~\ref{fig:ResidualsScatterplot}.

\textbf{Multicollinearity.} Finally, we checked for multicollinearity; that is, we evaluated whether the exogenous variables are correlated with each other. 
This step is essential because our primary goal is to isolate the relationship between the endogenous variable and each of the exogenous variables.
If the exogenous variables are correlated, changes in one of them will also influence another, leading to unreliable results.
A standard measure for assessing multicollinearity is the variance inflation factor (VIF). A commonly accepted cut-off for the VIF is 5, but a more conservative cut-off is 3 \citep{Miles2014VIF,Obrien2007VIF}.
Table~\ref{tab:VIF} shows that all VIF values are below two, well below the cut-off value.
Hence, we are confident that our regression model will associate the variance of the endogenous variable with the exogenous ones.

\begin{table}[!t]
	\sisetup{
    group-digits=true,
    group-minimum-digits=4,
    table-format=0.3,
    mode=text,
    detect-weight=true, 
    detect-family=true
}
	\caption{Coefficients for need for cognition to assess multicollinearity}
	\label{tab:VIF}
	{
		\begin{tabular}{p{4.5cm}SSSSS}
			\toprule
			Variables & {Unstandardized}	 & {Standardized} & {\textit{t}} & {\textit{p}} & {VIF}  \\
			\midrule
			Machiavellianism & 0.058 & 0.069 & 1.422 & 0.156 & 1.662  \\
			Psychopathy & 0.073 & 0.079 & 1.608 & 0.109 & 1.718  \\
			Narcissism & 0.012 & 0.014 & 0.318 & 0.751 & 1.400  \\
			Honesty-Humility & 0.258 & 0.260 & 5.226 & 0.000 & 1.747  \\
			Emotionality & -0.129 & -0.128 & -3.088 & 0.002 & 1.210  \\
			Extraversion & 0.060 & 0.060 & 1.384 & 0.167 & 1.316  \\
		    Agreeableness  & -0.097 & -0.082 & -1.958 & 0.051 & 1.235  \\
			Conscientiousness  & 0.237 & 0.217 & 5.268 & 0.000 & 1.202\\
		    Openness to Experiences & 0.426 & 0.360 & 8.976 & 0.000 & 1.135  \\
			\bottomrule
		\end{tabular}
	}
\end{table}

Another key step in a Bayesian analysis is the specification of a prior \citep{JASP2019guidelines,van2021tutorial,furia2021}. In our case, the prior of a Bayesian multi-model regression is composed of two parts. A first part refers to the prior over the 512 different linear regression models explored in this analysis. In our case, we decided to use a uniform prior (i.e. all models have the same prior probability). Eventually, a non-uniform prior could be employed encoding some previously available information about the relevance of some of our personality traits for predicting need for cognition. However, we did not find a sensible way to define a precise probability value for all (or some of) the personality traits reflecting the information present in the literature.  

The second part of the prior in a Bayesian multi-model regression refers to the prior over the linear coefficients associated to each of the independent variables (i.e. the personality traits). For example, a negative linear coefficient for a given personality trait indicates an negative correlation between this personality trait and need for cognition. Finding a prior for these continuous values is not as easy as in the case of the different models. Following the guidelines by \cite{JASP2019guidelines} and \cite{van2021tutorial}, we considered four different priors which are commonly used in the Bayesian literature for the coefficients of a linear regression model \citep{liang2008mixtures,rouder2012default}.
By using different priors, we can assess the sensitivity of our analysis to the choice of the prior, which is a standard robustness check in Bayesian data analysis \citep{JASP2019guidelines,van2021tutorial}. 
In the following section, we  report the average plus/minus the standard deviation of the different computations, such as the Bayes Factors and the posterior probabilities, obtained by repeating the whole data analysis using a different prior at a time. The standard deviations  will help us to identify the sensitivity of the Bayesian analysis to the choice of the prior. Small standard deviation values, in comparison to the mean value, 
suggests that the results do not vary much across different priors, and thus that the prior of choice does not have an impact on the results that would raise any concerns as to the sensitivity of the prior of choice. On the other hand, large standard deviation values, in comparison to the mean value, suggest that the choice of prior \textit{does} have an impact on the results, and this should be reason for concern, because the results depend on a specific prior.

The results of the various computations are included in the supplementary material at the end of this article.
In Appendix \ref{app:BLR}, we provide full details about these priors used and how to reproduced the analysis shown in this paper using JASP.

\subsection{Results}

We now present the results of our analysis. We establish (i) which personality traits best predict developers' need for cognition; (ii) how much developers' personality traits contribute in explaining their need for cognition; and (iii) whether or not software engineers have a higher need for cognition compared to a general population.

The result of the ten best regression models out the 512 models considered in this Bayesian analysis are summarized in Table~\ref{tab:modelComparison-NfC}.
The table shows P(H\textpipe{}data) 
which is the posterior probability of each model/hypothesis after observing the data, the Bayes factor (BF\textsubscript{10}),  
and the explained variance \textit{R}\textsuperscript{2}.
As noted, Table~\ref{tab:bayesian} offers heuristics to interpret Bayes factors. We  note that JASP fixes as the reference model (H\textsubscript{0} following previous notation) the model on top of the list (this is because BF\textsubscript{10} is equal to one in this case). The other models have Bayes factors smaller than one, showing that they are all less supported by the data compared to the reference model. Nevertheless, as can be seen, no model has an overwhelming advantage over the others. This suggests that any of these subsets of personality traits explains NfC similarly well. 
Based on the \textit{R}\textsuperscript{2} values, we conclude that personality traits explain approximately one-third of the variation in need for cognition of software engineers in our sample.

As discussed in the previous section, Table~\ref{tab:modelComparison-NfC} reports the average value plus/minus the standard deviation of the different quantities over four analysis, each of them performed with a different prior. As previously discussed, the standard deviations help us to identify the sensitivity of the Bayesian analysis to the choice of the prior. As can be seen, standard deviations values are very small and, in consequence, the main conclusions of this analysis are not affected by the particular choice of the prior.

\begin{table*}[!t]
\caption{Model Comparison --- need for cognition (machiavellianism (Mac), psychopathy (Psy), narcissism (Nar), honesty-humility (H-H), emotionality (Emo), extraversion (Ext), Agreeableness (Agr), conscientiousness (Con), openness to experiences (OtE)). We report the average value $\pm$ the standard deviation of the different Bayesian quantities over four analysis, each of them performed with a different prior, as detailed in Section \ref{sec:BMMLR}.}
\label{tab:modelComparison-NfC}
	
	\robustify\bfseries
	\sisetup{
    group-digits=true,
    group-minimum-digits=4,
    table-format=0.3,
    mode=text,
    detect-weight=true, 
    detect-family=true
}

	{
 		\begin{tabular}{clccc}
			\toprule
 			No. & Models & {P(H\textpipe{}data) } & {BF\textsubscript{10}} & {\textit{R}\textsuperscript{2}}  \\
			\cmidrule[0.4pt]{1-5}
1 & Mac + H-H + Emo + Agr + Con + OtE & 0.13$\pm$ 0.00 & 1.00$\pm$ 0.00 & 0.33 \\
2 & H-H + Emo + Agr + Con + OtE & 0.10$\pm$ 0.01 & 0.77$\pm$ 0.08 & 0.32 \\
3 & Psy + H-H + Emo + Con + OtE & 0.09$\pm$ 0.01 & 0.68$\pm$ 0.07 & 0.32 \\
4 & Psy + H-H + Emo + Agr + Con + OtE & 0.08$\pm$ 0.00 & 0.56$\pm$ 0.00 & 0.32 \\
5 & Mac + Psy + H-H + Emo + Con + OtE & 0.07$\pm$ 0.00 & 0.50$\pm$ 0.00 & 0.32 \\
6 & Mac + H-H + Emo + Con + OtE & 0.06$\pm$ 0.01 & 0.45$\pm$ 0.05 & 0.32 \\
7 & Mac + Psy + H-H + Emo + Agr + Con + OtE & 0.05$\pm$ 0.00 & 0.36$\pm$ 0.03 & 0.33 \\
8 & Psy + H-H + Emo + Ext + Agr + Con + OtE & 0.05$\pm$ 0.00 & 0.34$\pm$ 0.03 & 0.33 \\
9 & Mac + H-H + Emo + Ext + Agr + Con + OtE & 0.04$\pm$ 0.00 & 0.28$\pm$ 0.03 & 0.33 \\
10 & Psy + H-H + Emo + Ext + Con + OtE & 0.03$\pm$ 0.00 & 0.25$\pm$ 0.00 & 0.32 \\
			\bottomrule
		\end{tabular}
	}
\end{table*}

If we jointly consider the different models shown in Table~\ref{tab:modelComparison-NfC}, several traits appear in most of them. Table~\ref{tab:posteriorSummariesOfCoefficients}  lists the individual probabilities of each trait to be relevant for predicting need for cognition. The inclusion probability of a given trait is computed by summing up (i.e., marginalizing) the posterior probabilities of all of the models where this trait is included (all the 512 evaluated models are considered here, not only the ten ones shown in Table~\ref{tab:modelComparison-NfC}). Whereas Table~\ref{tab:modelComparison-NfC} considers traits only in \textit{combinations}, 
Table~\ref{tab:posteriorSummariesOfCoefficients} presents information on the likelihood of individual traits to be needed to predict the developer's need for cognition. Note that, as in the previous table, we report average values plus/minus the standard deviation in order to evaluate the sensitivity of the choice of the prior. 
%



\begin{table}[!ht]
	\sisetup{
    group-digits=true,
    group-minimum-digits=4,
    table-format=0.3,
    mode=text,
    detect-weight=true, 
    detect-family=true
}
	\caption{Posterior summaries of coefficients. {P(incl\textpipe{}data)} denotes the posterior inclusion probability. The last two columns denote the 95\% central credible interval (CI) for the value of the associated linear coefficients. We report the average value $\pm$ the standard deviation of the different quantities over four analyses, each of them performed with a different prior, as detailed in Section~\ref{sec:BMMLR}.}
	\label{tab:posteriorSummariesOfCoefficients}
	{
	
 		\begin{tabular}{lccc}
			\toprule
 			\multicolumn{1}{c}{}  & \multicolumn{1}{c}{} & \multicolumn{2}{c}{95\% CI} \\
 			\cmidrule{3-4}
			Coefficient & {P(incl\textpipe{}data)} & {Lower} & {Upper}  \\
			\cmidrule[0.3pt]{1-4}
Intercept & 1.00$\pm$ 0.0 & 3.92$\pm$ 0.0 & 4.02$\pm$ 0.0 \\
Mac & 0.49$\pm$ 0.0  & -0.00$\pm$ 0.0 & 0.13$\pm$ 0.0 \\
Psy & 0.52$\pm$ 0.0  & -0.00$\pm$ 0.0 & 0.16$\pm$ 0.0 \\
Nar & 0.16$\pm$ 0.0  & -0.02$\pm$ 0.0 & 0.06$\pm$ 0.0 \\
H-H & 1.00$\pm$ 0.0  & 0.13$\pm$ 0.0 & 0.32$\pm$ 0.0 \\
Emo & 0.96$\pm$ 0.0  & -0.22$\pm$ 0.0 & -0.02$\pm$ 0.0 \\
Ext & 0.26$\pm$ 0.0  & -0.01$\pm$ 0.0 & 0.10$\pm$ 0.0 \\
Agr & 0.61$\pm$ 0.0  & -0.18$\pm$ 0.0 & 0.00$\pm$ 0.0 \\
Con & 1.00$\pm$ 0.0  & 0.15$\pm$ 0.0 & 0.32$\pm$ 0.0 \\
OtE & 1.00$\pm$ 0.0  & 0.34$\pm$ 0.0 & 0.52$\pm$ 0.0 \\
            \bottomrule
		\end{tabular}
	}
\end{table}

As a result of this Bayesian analysis, four traits in particular have an extremely high chance of almost 100\% to be needed to predict need for cognition, namely: openness to experience (OtE), conscientiousness (Con), honesty-humility (H-H), and emotionality (Emo). This is reflected in the probability of inclusion in the regression model after observing the data (i.e., P(incl\textpipe{}data)), with values of  100\% (with the only exception being emotionality which has a probability of 96.3\%). Note that in this analysis, we assumed that \textit{a priori} (i.e., before considering the data from the survey), all traits had the same chance of 50\% to be relevant for predicting need for cognition (i.e., P(incl)=0.5).



The next question is whether these personality traits are positively, negatively, or not correlated with the need for cognition. Table~\ref{tab:bayesianPearsonCorrelations} presents the Pearson coefficients that represent the relations between need for cognition and personality traits. Based on the correlation using Pearson's \textit{r}, we can see the sign (negative or positive) of the individual relation but not the strength of the evidence supporting this relationship. To address this issue, we also compute the Bayes factor for a model where no relationship is established between the trait and need for cognition. In this case, the specification of the prior is different from the case of the Bayesian multi-model regression because we now use a simple correlation model. Again, we consider four different priors and report average plus/minus standard deviations for the quantities depending on the prior. In this case, the standard deviations of some of the Bayes factors are, at first sight, relatively high, but we should note that in these cases, Bayes factors always vary within a range of extremely supportive evidence according to the ranges provided in Table~\ref{tab:bayesian}. This effect can be appreciated when looking at the associated posterior probabilities, which can be computed from the Bayes factors, because, in those cases, the associated standard deviations become negligible. Again, we can see that the main conclusions of this analysis are not affected by the particular choice of the prior. In Appendix \ref{app:BLR}, we provide more details about these priors and about how to reproduce this analysis using JASP.

In particular, we can see how dark traits (i.e., psychopathy, machiavellianism, and narcissism) are negatively associated with need for cognition, although those associations are barely significant (i.e., they have negative Pearson's r coefficients but very small Bayes factors). Honesty-humility, extraversion, conscientiousness, and openness to experience are significantly positively associated with need for cognition (i.e., they have positive Pearson's r coefficients and large Bayes factors). Also, emotionality is strongly negatively associated to NfC. However, according to this analysis, we can not infer whether agreeableness is associated to NfC in a significant way.


	

\begin{table}[!ht]
	\caption{Bayesian correlation analysis between need for cognition and personality traits. We report the average value $\pm$ the standard deviation of the different Bayesian quantities over four analysis, each of them performed with a different prior, as detailed in Appendix \ref{app:BLR}.}
	\label{tab:bayesianPearsonCorrelations}
		\sisetup{
    group-digits=true,
    group-separator={,},
    group-minimum-digits=4,
    table-format=-1.3, 
    mode=text,
    detect-weight=true, 
    detect-family=true
}
		\begin{tabular}{lScc}
			\toprule
			Trait & {Pearson's \textit{r}} & {P(correlation\textpipe{}data)} & {BF\textsubscript{10}}\\
			\midrule
Mac & -0.02 & 0.06$\pm$ 0.02 & 0.06$\pm$ 0.02 \\ 
Psy & -0.04 & 0.07$\pm$ 0.02 & 0.08$\pm$ 0.03 \\ 
Nar & -0.07 & 0.17$\pm$ 0.05 & 0.21$\pm$ 0.07 \\ 
H-H & 0.24 & 1.00$\pm$ 0.00 & 4.3e+04$\pm$ 1.3e+04 \\ 
Emo & -0.20 & 1.00$\pm$ 0.00 & 1.1e+03$\pm$ 3.3e+02 \\ 
Ext & 0.20 & 1.00$\pm$ 0.00 & 1.2e+03$\pm$ 3.6e+02 \\ 
Agr & -0.11 & 0.44$\pm$ 0.08 & 0.81$\pm$ 0.27 \\ 
Con & 0.35 & 1.00$\pm$ 0.00 & 3.3e+12$\pm$ 8.3e+11 \\ 
OtE & 0.42 & 1.00$\pm$ 0.00 & 2.2e+19$\pm$ 4.6e+18 \\ 
 			\bottomrule
 			\end{tabular}

\end{table}

\begin{figure}[!ht]
\includegraphics[trim={0cm 21mm 0cm 0cm },clip,width=3in]{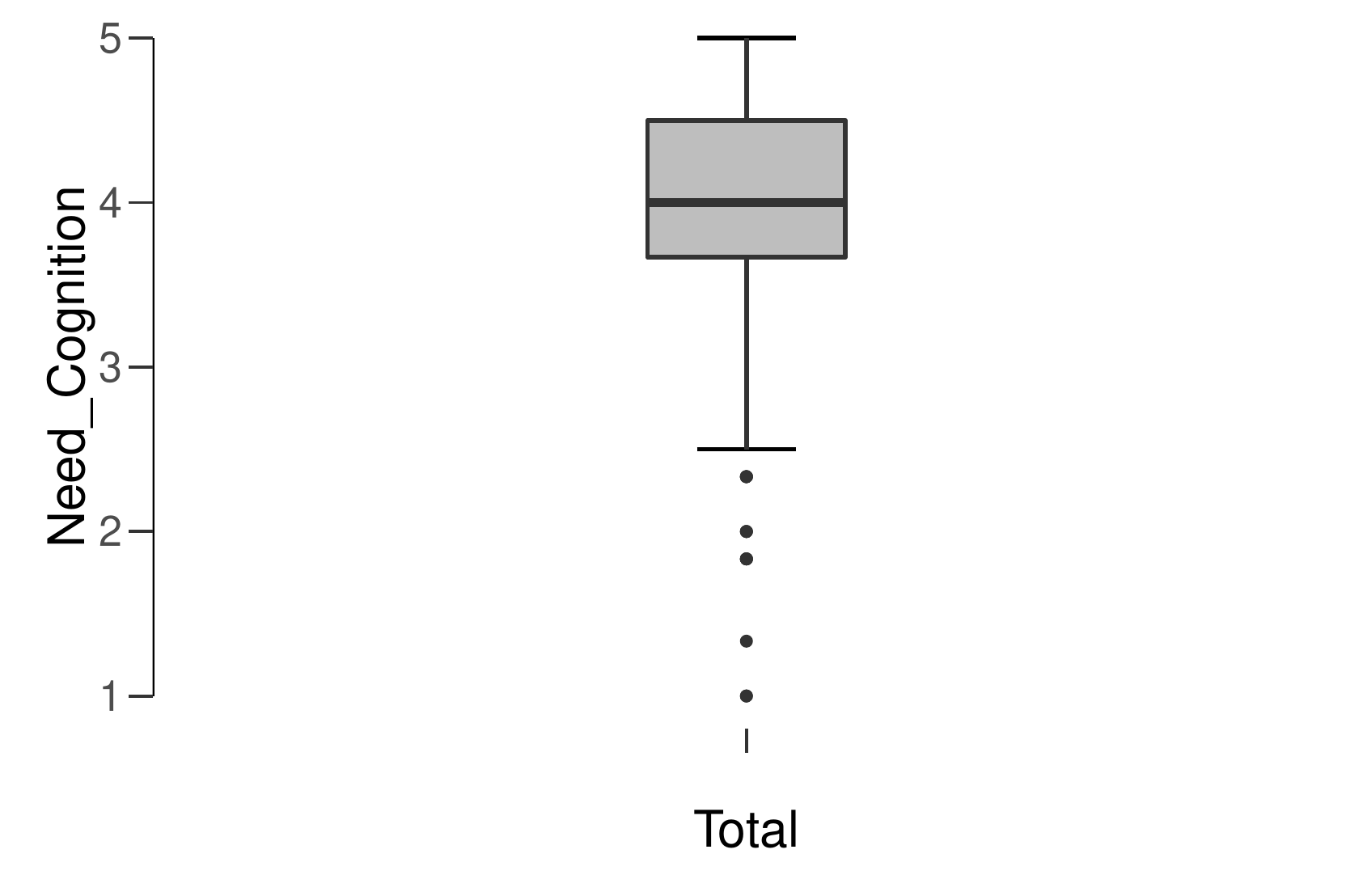}
\caption{Distribution of software professionals' need for cognition.}
\label{fig:NfC}
\end{figure}

Finally, we assess the level of need for cognition for software engineers through an analysis of variance (ANOVA).
To do so, we consider the descriptive statistics in Table~\ref{tab:NfC_DescriptiveStatistics} and the boxplot in Figure~\ref{fig:NfC}.
To interpret these values, we compare these to findings by  \cite{popoviciu2011students}, who observed  a mean value of 3.97 (on a scale of 1-5), suggesting that software professionals are ``very likely to engage in and enjoy effortful cognitive activities.''

\begin{table}[!ht]
	\sisetup{
    group-digits=true,
    group-minimum-digits=4,
    table-format=0.3,
    mode=text,
    detect-weight=true, 
    detect-family=true
}
	\caption{Descriptive statistics of need for cognition}
	\label{tab:NfC_DescriptiveStatistics}
	{
		\begin{tabular}{lS}
			\toprule
			Statistic & {Value} \\
			\cmidrule[0.4pt]{1-2}
			N & 483  \\
        	Minimum value & 1  \\
			Maximum value & 5  \\
			Mean & 3.970  \\
			Standard Error of Mean & 0.032  \\
			Standard Deviation & 0.694  \\
			Skewness & -0.926  \\
			Standard Error of Skewness & 0.111  \\
			Kurtosis & 1.643  \\
			Standard Error of Kurtosis & 0.222  \\
			\bottomrule
		\end{tabular}
	}
\end{table}

\begin{table}[!ht]
	\sisetup{
    group-digits=true,
    group-minimum-digits=4,
    table-format=0.3,
    mode=text,
    detect-weight=true, 
    detect-family=true
}
	\caption{Need for cognition scores in this and other studies} 
	\label{tab:NfC_genpop}
	{
		\begin{tabular}{llSS}
			\toprule
			Study & {Sample size} & {Mean} & {SD} \\
			\midrule
			\cite{kearney2009and} & 83 teams (549 professionals) & 3.33 & 0.22  \\
        	\cite{popoviciu2011students} & 30 graduate students & 3.56 & {NA} \\
        	\cite{madrid2016creativity} & 220 professionals & 3.79 & 0.60 \\
	This study  & 483 software engineers & 3.97 & 0.69 \\
			\bottomrule
		\end{tabular}
	}
\end{table}

We now compare those values with previous studies reporting scores of need for cognition, summarized in Table~\ref{tab:NfC_genpop}.
Based on the reported mean and standard deviation (SD) values of need for cognition, as reported by \cite{kearney2009and} and  \cite{madrid2016creativity}, we replicate the prior research data using two normal distributions (we could not replicate \cite{popoviciu2011students}, since the standard deviation was not reported). We subsequently use these distributions to compare prior work to our results.
First, we ran a one-way ANOVA comparing need for cognition between these two studies and our results. This shows a significant effect of the study sample on need for cognition (F (2) = 38.16, \textit{p} \textless{} 0.01).
The ANOVA confirms that there is an overall significant difference among groups; however, it does not provide any information about which group scores significantly higher in need for cognition.
Therefore, we performed a two-way Tukey posthoc comparison between the studies.
This test is a widespread posthoc analysis based on the assumption of normally distributed data. 

\begin{figure}[!ht]
\includegraphics[width=3.8in]{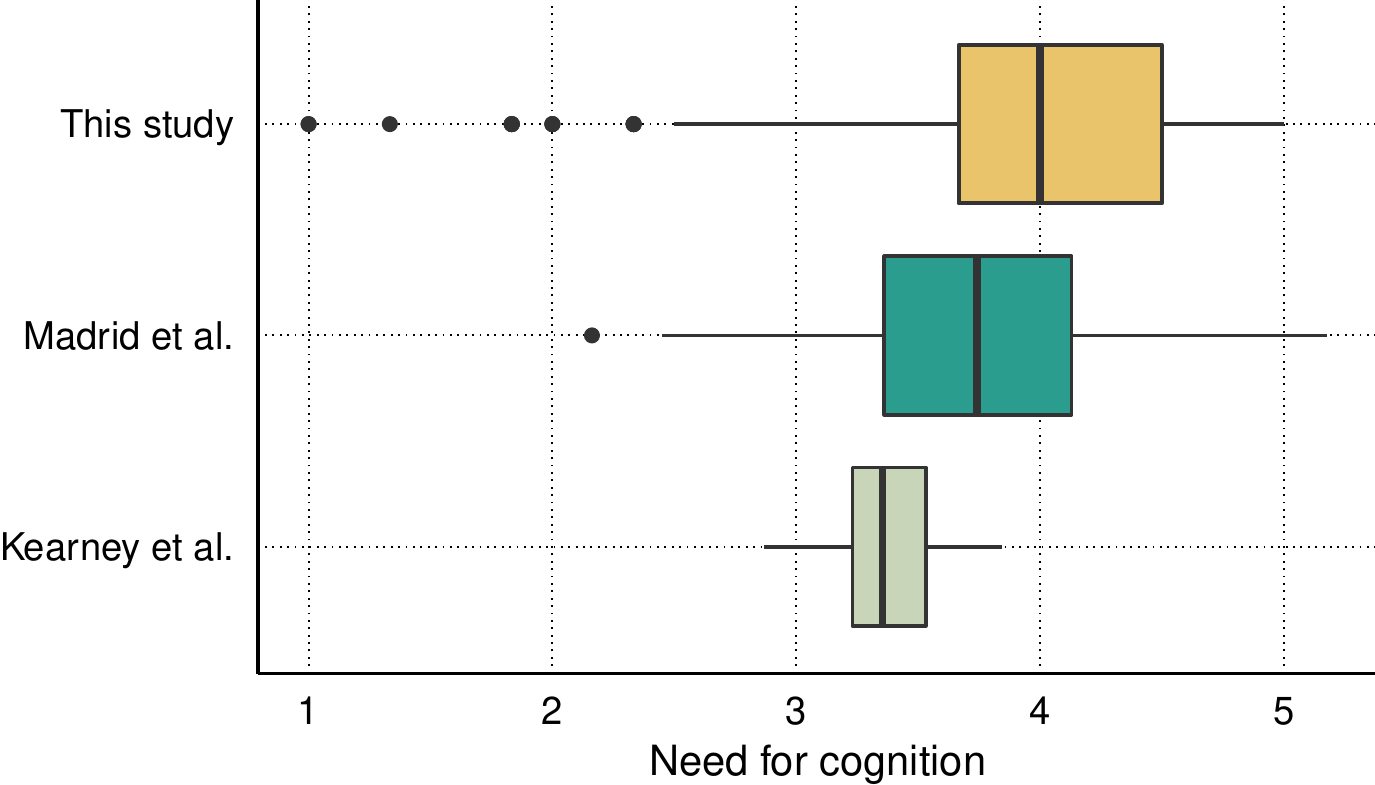}
\caption{Boxplot distribution of need for cognition studies}
\label{fig:Boxplot}
\end{figure}

This analysis shows that need for cognition in our study is significantly higher, compared to \cite{kearney2009and} (\textit{p} \textless{} 0.001) and \cite{madrid2016creativity} (\textit{p} = 0.001).
Figure~\ref{fig:Boxplot} shows the boxplot distribution for need for cognition scores. 
We conclude, thus, that software engineers report a higher need for cognition compared to other professionals.

Upon closer inspection of need for cognition, 
Table~\ref{tab:NfC_DescriptiveStatistics} shows that the mean value is high, close to four (3.97) on a scale of one to five.
Also, the descriptive statistics of the variables suggest that need for cognition is relatively skewed towards high values.
The boxplot shown in Figure~\ref{fig:NfC} confirms this, where both upper and lower quartiles are close to the median value of four, with very few outliers below the lower whisker.
Accordingly, we conclude that software engineers have a significantly higher need for cognition than other knowledge workers.

\section{Discussion}
\label{sec:discussion}

\subsection{Discussion of Findings}
Based on the results of our analysis presented in the previous section, we can now address our central research question: What is the relationship between the personality traits of software engineers and their need for cognition?

Following increasing attention to human factors in general and personality traits in particular, this paper presents an investigation that links personality traits and the need for cognition among software professionals. 
Psychology literature suggests that personality is stable over time; i.e., it does not overall change along the maturation process of people \citep{cobb2012stability,gustavsson1997stability,blonigen2008stability}.
This stability of personality has also been confirmed for software developers \citep{calefato2019large}.
Need for cognition has also been identified as a stable psychological attribute \citep{furnham2013stable,roberts2008stable}.
A longitudinal study confirmed this stability over time\footnote{For the sake of completeness, need for cognition slightly increases over time for young individuals and slightly decreases for older ones (in a statistically non-significant way). However, Bruinsma and Crutzen did not consider confounding factors that might explain these trends, such as dementia which could mostly explain such decrease.} \citep{bruinsma2018longitudinal}.
This evidence allows us to make reliable predictions because all traits included in this study are stable and not time-dependent. 
Therefore, we can conclude that software engineers have a higher need for cognition for the general population, which is predicted by their defining characteristics, i.e., personality traits.
Although we did not investigate causality, we found that personality traits are good predictors of developers' high need for cognition. 
Moreover, since both bright and dark personality traits and the need for cognition are stable over time, it suggests that the predictive relationship is also stable over time.

Anecdotal evidence has suggested that software engineers must enjoy effortful thinking to be good developers \citep{petre1997}.
This investigation aims to move from anecdote to evidence through a large-scale empirical study.
Operationalizing the joy of thinking with the need for cognition (as shown by  \cite{strobel2018predicting}) brings several advantages to the community's body of knowledge because it allows us to infer previous evidence on the need for cognition with tailored recommendations for software professionals. 

We now focus our discussion on the regression model and the role of each trait.
From the posteriors of Table~\ref{tab:posteriorSummariesOfCoefficients}, we can see that four traits have a probability of inclusion that is close to 100\%, namely openness to experience, conscientiousness, honesty-humility, and emotionality. 
These are the only four traits that have a significant BF\textsubscript{inclusion} (i.e., higher than ten \citep{lee2014bayesian}), and are present in all ten of the most likely regression models listed in Table~\ref{tab:modelComparison-NfC}.
To recall, we reported only the ten best models of the Bayesian regression that showed the greater likelihood to predict our dependent variable (i.e., Need for Cognition).
In particular, based on the correlation signs of Table~\ref{tab:bayesianPearsonCorrelations}, \textit{we conclude that high openness to experience, high conscientiousness, high honesty-humility, and low emotionality predict a high level of need for cognition.}
Similar patterns have also been observed in a previous study \citep{sadowski1997need}. 

Prior studies of personality traits in software engineering are limited to only bright personality traits \citep{cruz2015forty}.
This study is among the first to include  dark personality traits.\footnote{In a previous study, we focused on gender differences in the same set of bright and dark personality traits of software engineers \citep{russo2020gender}.}
The inclusion of dark traits offers a more comprehensive and holistic understanding of personality traits of software engineers and, in this study, in particular, how these relate to developers' need for cognition.
However, Table~\ref{tab:bayesianPearsonCorrelations} shows that none of the three dark traits, machiavellianism, psychopathy, and narcissism, are significantly correlated to the need for cognition (with a BF\textsubscript{10} close to 0).
We therefore conclude that bright traits play a much greater role in predicting the need for cognition than dark traits.
Furthermore, the BF\textsubscript{inclusion} values of these dark traits are also relatively low.
This suggests that dark traits do not play a significant role in the need for the cognition of software professionals.

Additionally, our ANOVA analysis confirmed that \textit{software professionals score significantly higher in their need for cognition compared to the general population.}
Need for cognition seems to be one unique characteristic of developers.

Finally, we also highlight that personality traits predict one-third of the  need for cognition software engineers.
This evidence has important implications for both practice and research discussed next.

\subsection{Implications for Research and Practice}

Table~\ref{tab:implications} summarizes the key findings of our study and provides recommendations for software organizations.
First, our study finds that software engineers score, on average, high on the construct need for cognition. 
Second, in seeking to predict need for cognition with personality traits, we find that the latter account for ca. 33\% of the variation in need for cognition. This is a crucial finding of this study and has several implications for research and practice. 

\hyphenation{experience}
\begin{table*}[!ht]
\caption{Summary of findings and propositions}
\label{tab:implications}
\begin{tabular}{p{1.7cm}p{3cm}p{6.3cm}}
\toprule
Theme & Findings & Propositions \\ \midrule
Prediction & Personality traits of software professionals are good predictors for their need for cognition. & Personality traits of software engineers are able to explain 33\% of the variation in need for cognition. \\ \addlinespace
Recruitment & Personality traits of software engineers predict a high need for cognition. & To attract software professionals, organizations should provide cognitive recruitment messages and not emotional ones. \\ 
\addlinespace
Teaming & Personality traits of software engineers predict a high need for cognition. & Software team composition with an individual high in need for cognition enhances task-relevant information and collaborative team identification, potentially leading to better team performance. \\
\addlinespace
Working behavior & Personality traits of software engineers predict a high need for cognition. & Software engineers have important characteristics that have to be embraced and reflected by a software organization. In particular, professionals tend to generate new ideas and are open to them. They are sensitive to organizational fairness and have a high work ethic. Also, individual innovation behavior is likely to be high. \\ 
\bottomrule
\end{tabular}
\end{table*}

Recruitment strategies can be better tailored for software engineers.
A study showed that affective (i.e., feeling-based) recruitment messages on company websites are the most attractive for a general population \citep{kraichy2014tailoring}.
However, for an individual with a high need for cognition, such messages are rather uninteresting. 
The same research showed that software professionals are more receptive to cognitive (i.e., thought-based) recruiting messages.
From a practical perspective, software organizations should not devote resources to create an emotional link between themselves and candidates.
Instead, it should focus on communicating fairly upon reasons why they are attractive to future employees.
One example is to describe in recruitment advertisements work‐life benefits since they positively influence applicant perceptions of the software company
\citep{firfiray2017lure}.

Moreover, regarding software teaming, the need for cognition predicts team performance \citep{kearney2009and}.
According to Kearney et al., a team with a high need for cognition can better process task-relevant information. 
For example, such a team will better understand the customer's needs of the technology stack they are supposed to work on.
Also, the collaborative team identification will be higher since the team cohesion will not depend on emotions but cognitive motivations. 
These two points are significant predictors of how a team can perform, i.e., delivering quality software.

Further, a high need for cognition of software engineers likely influences also working behaviors.
The working ethic of an individual with a high need for cognition is above-average \citep{singer1998consideration}.
Research by Singer et al. highlighted employees' high level of empathy with a high need for cognition with potential victims (i.e., colleagues who have been unjustly treated) of their organizations, which predicted the likelihood of whistleblowing behavior and harmful consequences for the company. 
Similarly, \cite{pohling2019moral} suggests that individuals' high need for cognition increases active cooperation and pro-social behaviors within organizations.
Thus, if developers' working environment is perceived as unfair, unwelcome, or toxic by developers, they will likely start looking for another workplace and increase their whistleblowing intention \citep{valentine2019moral}.
Consequently, an organizational culture based on fairness, transparency, and accountability is considered very important for subjects high in need of cognition, such as software professionals.
Accordingly, software companies that embrace such organizational behaviors will more likely experience a higher sense of belonging and community of their employees.

Likewise, although individuals such as software engineers are open to change, they do not follow it without questioning it \citep{haugtvedt1992personality}. 
They tend to have a critical approach, especially towards contradictory or ill-defined problems.
Therefore, whatever changes a software organization might be willing to introduce, it must ensure that such changes have been communicated sufficiently precisely so that software engineers understand the need and benefits of such change; otherwise, they will be reluctant to change stick to old habits. 

Another appealing implication to bear in mind is a high degree of creativity at work \citep{madrid2016creativity}.
Cognitive-laden individuals are very likely to generate new ideas.
Interestingly, as \cite{madrid2016creativity} found, a fair working environment leads to novel idea implementation. 
These findings are essential in creative jobs, where employees have to develop new solutions, such as software development. 
A fair and transparent organizational culture plays an essential role in delivering customers with novel solutions.
Similarly, Wu et al. reported that the need for cognition is an antecedent of individual innovation behaviors.
In other words, the authors suggest that the high need for cognition explains innovation behaviors.
This results in a higher degree of job autonomy and the ability to better handle time pressure \citep{wu2014need}.
It is also worth mentioning that people scoring high on need for cognition are less prone to the so-called primacy effect: they recall presented information comprehensively and uniformly and are not biased by the first items of a list, showing a higher mental flexibility~\citep{ahlering1989need}.

\subsection{Threats to Validity}
We now discuss threats to validity following \cite{gren2018standards}'s validity concerns for behavioral software engineering studies. 

\textit{Reliability}. 
We only used stable and reliable measurement instruments that have been validated by social psychology scholars and used in a large number of prior studies. For example, HEXACO has been used in almost 1,000 studies.\footnote{www.hexaco.org/reference}

\textit{Construct validity}. 
Our three measurement instruments reflect the constructs for which they have been developed.
Constructs are well-established in the scholarly debate.
Instruments have been refined and improved over time. 
We acknowledge having used short versions of the measurement instruments, which were slightly less accurate.
Nonetheless, all three instruments can estimate with high accuracy more extended inventories.
This permitted us to be parsimonious and to use multiple psychometric models (i.e., bright, dark traits, and the need for the cognition of software developers), with a high engagement rate and a low drop-out ratio in the study.

\textit{Conclusion validity}.
Bayesian statistics help us to gather a more comprehensive understanding of the subject matter, overcoming the typical shortcomings of \textit{p}-values-based findings of frequentist null-hypothesis significance testing. 
Conclusions are drawn based on the odds related to the best predictive model and not just by answering yes or no to a specific hypothesis.
We also stress the fact that we did not observe the discussed implications directly.
Instead, we inferred our findings with relevant literature on the subject matter.
Accordingly, we developed four propositions that need further investigation.
We artificially generated normal distributions of previous studies to compute the ANOVA since we did not have access to such studies'  data for replication.

\textit{External validity}.
External validity is a crucial concern when conducting sample studies, which seek to maximize the potential for generalizability \citep{stol2018abc}. 
In order to establish a sample size that is representative of the software developer population, we relied on \cite{yamane1973statistics}'s formula. 
While there are several estimates of the size of the global software developer pool \citep{russo2020gender}, we used a conservative (i.e., the highest) estimate, which was approximately 36.5 million at the time of our study. Following Yamane's formula, a sample of ca. 400 was suggested. Our sample is well above that minimum.
We put the highest care in our multi-stage data collection process using a cluster-randomized sampling strategy, through which we gathered 483 validated responses.
Such a multi-stage process aimed to identify software professionals in our cluster.
As a cluster, we used the academic crowdsourcing platform Prolific.
Although cluster-randomized sampling is considered less efficient than standard-randomized sampling, it provides a fair representation of a population and is very cost-effective \citep{gravetter2018research}.
This approach also follows the most recent recommendations for sampling strategies in software engineering research \citep{baltes2020sampling}.

\section{Conclusion}
\label{sec:conclusion}

The individual motivation to pursue a software engineering career is a puzzling issue to address since there are so many confounding factors that are very hard to control, e.g., the socio-economic background, individual experiences, and aspirations.
Nevertheless, some specific psychological characteristics are intuitively linked to any software professional, such as the joy for problem-solving and, more generally, thinking. 
Therefore, we used in this study need for cognition (i.e., a tendency to engage and enjoy effortful thinking \citep{cacioppo1982need}) as a regressor of individual defining or innate characteristics, namely personality traits.
Our analysis shows that the personality traits of software engineers have a considerable explanatory power with a coefficient of determination (\textit{R}\textsuperscript{2}) of ca. 0.33 of the need for cognition.
Thus, individuals who have a high need for cognition are more likely to start a career in software engineering than those who do not.
This finding also has several managerial propositions for software organizations since software professionals are analytical subjects, likely prefer cognitive communication with a transparent, accountable, and fair organizational culture.

Most of the insights we have reported might be seen as self-evident or already apparent, at least to some practitioners.
Nevertheless, we wish to stress that so far, those insights had no scholarly grounding.
Therefore, we put considerable effort into recruiting a large sample of almost 500 software professionals to claim generalization of results, who have been screened throughout a multi-stage selection process, using a cluster-randomized sampling strategy.
Data have been analyzed through Bayesian linear regression to provide a solid understanding of the studied phenomenon.

This is regression analysis, and it does not assess any causal relationship between independent variables and dependent opens.
Nevertheless, our research design implicitly assumes a causal-predictive assumption of our theoretical hypothesis that has been validated.
Also, we observe that the general population scores significantly lower than software engineers in need of cognition and that the personality traits of developers can predict the individual need for cognition.
Accordingly, we speculate that people with a high need for cognition are more likely to start a career in computer science than those who score low.
In other words, there are some individual characteristic that will lead someone more naturally to a computer science career compared to others.
This insight is, for now, an educated guess and needs to be carefully validated with a proper research design to assess causation.

Although we grounded our recommendations in well-established social psychology and management literature, we have not reported them directly with this study.
Therefore, further research should investigate the suggested relations in a software engineering population.

\section*{Supplementary Material}
Raw data, survey questions, and additional material are openly available under a CC-BY 4.0 license at the following URL:
\url{https://osf.io/n57as/}.

\section*{Acknowledgments}
This work was supported by the Science Foundation Ireland grant 13/RC/2094\_P2 and 15/SIRG/3293 and co-funded under the European Regional Development Fund through the Southern \& Eastern Regional Operational Programme to Lero---the Irish Software Research Centre (www.lero.ie). For the purpose of Open Access, the authors have applied a CC-BY public copyright licence to any Author Accepted Manuscript version arising from this submission.

\section*{Conflict of Interest}
All authors declare that they have no conflicts of interest.



\bibliographystyle{spbasic}      

\bibliography{bibliography}

\begin{thebibliography}{125}
\providecommand{\natexlab}[1]{#1}
\providecommand{\url}[1]{{#1}}
\providecommand{\urlprefix}{URL }
\expandafter\ifx\csname urlstyle\endcsname\relax
  \providecommand{\doi}[1]{DOI~\discretionary{}{}{}#1}\else
  \providecommand{\doi}{DOI~\discretionary{}{}{}\begingroup
  \urlstyle{rm}\Url}\fi
\providecommand{\eprint}[2][]{\url{#2}}

\bibitem[{Ahlering and Parker(1989)}]{ahlering1989need}
Ahlering RF, Parker LD (1989) Need for cognition as a moderator of the primacy
  effect. Journal of Research in Personality 23(3):313--317

\bibitem[{Almstrum(2003)}]{almstrum2003attraction}
Almstrum VL (2003) What is the attraction to computing? Communications of the
  ACM 46(9):51--55

\bibitem[{Amichai-Hamburger et~al(2007)Amichai-Hamburger, Kaynar, and
  Fine}]{amichai2007effects}
Amichai-Hamburger Y, Kaynar O, Fine A (2007) The effects of need for cognition
  on internet use. Computers in Human Behavior 23(1):880--891

\bibitem[{Ashton(2013)}]{ashton2013individual}
Ashton MC (2013) Individual differences and personality. Academic Press

\bibitem[{Ashton et~al(2004)Ashton, Lee, Perugini, Szarota, De~Vries, Di~Blas,
  Boies, and De~Raad}]{ashton2004six}
Ashton MC, Lee K, Perugini M, Szarota P, De~Vries RE, Di~Blas L, Boies K,
  De~Raad B (2004) A six-factor structure of personality-descriptive
  adjectives: solutions from psycholexical studies in seven languages. Journal
  of Personality and Social Psychology 86(2):356

\bibitem[{Ashton et~al(2014)Ashton, Lee, and De~Vries}]{ashton2014hexaco}
Ashton MC, Lee K, De~Vries RE (2014) The {HEXACO} honesty-humility,
  agreeableness, and emotionality factors: A review of research and theory.
  Personality and Social Psychology Review 18(2):139--152

\bibitem[{Baker(2016)}]{baker2016statisticians}
Baker M (2016) Statisticians issue warning over misuse of p values. Nature
  531(7593):151

\bibitem[{Baltes and Ralph(2020)}]{baltes2020sampling}
Baltes S, Ralph P (2020) Sampling in software engineering research: A critical
  review and guidelines. arXiv

\bibitem[{Beecham et~al(2008)Beecham, Baddoo, Hall, Robinson, and
  Sharp}]{beecham2008motivation}
Beecham S, Baddoo N, Hall T, Robinson H, Sharp H (2008) Motivation in software
  engineering: A systematic literature review. Information and Software
  Technology 50(9-10):860--878

\bibitem[{Bergande et~al(2020)Bergande, Petrikoglou, Kaskalis, and
  Brune}]{bergande2020codetripping}
Bergande B, Petrikoglou A, Kaskalis TH, Brune P (2020) Codetripping: Towards
  mastering app development using a web-based learning tool. a case study. In:
  Proceedings of the European Conference on Software Engineering Education, pp
  47--51

\bibitem[{van~den {B}ergh et~al(2021)van~den {B}ergh, Clyde, Gupta, de~Jong,
  Gronau, Marsman, Ly, and Wagenmakers}]{van2021tutorial}
van~den {B}ergh D, Clyde MA, Gupta ARKN, de~Jong T, Gronau QF, Marsman M, Ly A,
  Wagenmakers EJ (2021) A tutorial on {B}ayesian multi-model linear regression
  with {BAS} and {JASP}. Behavior research methods pp 1--21

\bibitem[{van Berkel et~al(2021)van Berkel, Goncalves, Russo, Hosio, and
  Skov}]{van2021effect}
van Berkel N, Goncalves J, Russo D, Hosio S, Skov MB (2021) Effect of
  information presentation on fairness perceptions of machine learning
  predictors. In: Proceedings of the 2021 CHI Conference on Human Factors in
  Computing Systems, pp 1--13

\bibitem[{Blonigen et~al(2008)Blonigen, Carlson, Hicks, Krueger, and
  Iacono}]{blonigen2008stability}
Blonigen DM, Carlson MD, Hicks BM, Krueger RF, Iacono WG (2008) Stability and
  change in personality traits from late adolescence to early adulthood: A
  longitudinal twin study. Journal of personality 76(2):229--266

\bibitem[{Boehm and Papaccio(1988)}]{boehm1988understanding}
Boehm BW, Papaccio PN (1988) Understanding and controlling software costs. IEEE
  Transactions on Software Engineering 14(10):1462--1477

\bibitem[{Boyle(1995)}]{boyle1995myers}
Boyle GJ (1995) Myers-briggs type indicator (mbti): Some psychometric
  limitations. Australian Psychologist 30(1):71--74

\bibitem[{Bruinsma and Crutzen(2018)}]{bruinsma2018longitudinal}
Bruinsma J, Crutzen R (2018) A longitudinal study on the stability of the need
  for cognition. Personality and Individual Differences 127:151--161

\bibitem[{Cacioppo and Petty(1982)}]{cacioppo1982need}
Cacioppo JT, Petty RE (1982) The need for cognition. Journal of Personality and
  Social Psychology 42(1):116

\bibitem[{Cacioppo et~al(1996)Cacioppo, Petty, Feinstein, and
  Jarvis}]{cacioppo1996dispositional}
Cacioppo JT, Petty RE, Feinstein JA, Jarvis WBG (1996) Dispositional
  differences in cognitive motivation: The life and times of individuals
  varying in need for cognition. Psychological Bulletin 119(2):197

\bibitem[{Calefato et~al(2019)Calefato, Lanubile, and
  Vasilescu}]{calefato2019large}
Calefato F, Lanubile F, Vasilescu B (2019) A large-scale, in-depth analysis of
  developers’ personalities in the apache ecosystem. Information and Software
  Technology

\bibitem[{Callan et~al(2017)Callan, Kim, Gheorghiu, and
  Matthews}]{callan2017interrelations}
Callan MJ, Kim H, Gheorghiu AI, Matthews WJ (2017) The interrelations between
  social class, personal relative deprivation, and prosociality. Social
  Psychological and Personality Science 8(6):660--669

\bibitem[{Capretz(2003)}]{capretz2003}
Capretz LF (2003) Personality types in software engineering. International
  Journal of Human-Computer Studies 58(2):207--214

\bibitem[{Cegielski and Hall(2006)}]{cegielski2006makes}
Cegielski CG, Hall DJ (2006) What makes a good programmer? Communications of
  the ACM 49(10):73--75

\bibitem[{Cobb-Clark and Schurer(2012)}]{cobb2012stability}
Cobb-Clark DA, Schurer S (2012) The stability of big-five personality traits.
  Economics Letters 115(1):11--15

\bibitem[{Cohen(1994)}]{cohen1994earth}
Cohen J (1994) The earth is round (p \textless{} . 05). American Psychologist
  49(12):997

\bibitem[{Corr and Matthews(2009)}]{corr2009personality}
Corr PJ, Matthews G (2009) The Cambridge handbook of personality psychology.
  Cambridge University Press Cambridge

\bibitem[{Cruz et~al(2015)Cruz, da~Silva, and Capretz}]{cruz2015forty}
Cruz S, da~Silva FQ, Capretz LF (2015) Forty years of research on personality
  in software engineering: A mapping study. Computers in Human Behavior
  46:94--113

\bibitem[{Cucolas and Russo(2021)}]{cucolas2021impact}
Cucolas AA, Russo D (2021) The impact of working from home on the success of
  scrum projects: A multi-method study. arXiv preprint arXiv:210705955

\bibitem[{Curtis(1984)}]{curtis1984fifteen}
Curtis B (1984) Fifteen years of psychology in software engineering: Individual
  differences and cognitive science. In: Proceedings of the 7th International
  Conference on Software Engineering, IEEE, pp 97--106

\bibitem[{Da~Cunha and Greathead(2007)}]{da2007does}
Da~Cunha AD, Greathead D (2007) Does personality matter? an analysis of
  code-review ability. Communications of the ACM 50(5):109--112

\bibitem[{De~Vries(2013)}]{de2013BHI}
De~Vries RE (2013) The 24-item brief {HEXACO} inventory ({BHI}). Journal of
  Research in Personality 47(6):871--880

\bibitem[{Devito(1985)}]{devito1985review}
Devito AJ (1985) Review of {M}yers-{B}riggs type indicator. The ninth mental
  measurements yearbook pp 1030--1032

\bibitem[{Dienes(2014)}]{dienes2014Bayes}
Dienes Z (2014) Using {B}ayes to get the most out of non-significant results.
  Frontiers in psychology 5:781

\bibitem[{van Doorn et~al(2021)van Doorn, van~den Bergh, B\"{o}hm, Dablander,
  Derks, Draws, Etz, Evans, Gronau, Haaf, Hinne, Kucharsk{\`y}, Ly, Marsman,
  Akash R. Komarlu Narendra Gupta~aand Sarafoglou, Stefan, Voelkel, and
  Wagenmakers}]{JASP2019guidelines}
van Doorn J, van~den Bergh D, B\"{o}hm U, Dablander F, Derks K, Draws T, Etz A,
  Evans NJ, Gronau QF, Haaf JM, Hinne M, Kucharsk{\`y} {\v{S}}, Ly A, Marsman D
  Maarten aand~Matzke, Akash R Komarlu Narendra Gupta~aand Sarafoglou A, Stefan
  A, Voelkel JG, Wagenmakers EJ (2021) The jasp guidelines for conducting and
  reporting a {B}ayesian analysis. Psychonomic Bulletin and Review
  28(3):813--826

\bibitem[{Druckman and Bjork(1991)}]{druckman1991mind}
Druckman DE, Bjork RA (1991) In the mind's eye: Enhancing human performance.
  National Academy Press

\bibitem[{Evans and Simkin(1989)}]{evans1989best}
Evans GE, Simkin MG (1989) What best predicts computer proficiency?
  Communications of the ACM 32(11):1322--1327

\bibitem[{Feldt et~al(2010)Feldt, Angelis, Torkar, and
  Samuelsson}]{feldt2010links}
Feldt R, Angelis L, Torkar R, Samuelsson M (2010) Links between the
  personalities, views and attitudes of software engineers. Information and
  Software Technology 52(6):611--624

\bibitem[{Firfiray and Mayo(2017)}]{firfiray2017lure}
Firfiray S, Mayo M (2017) The lure of work-life benefits: Perceived
  person-organization fit as a mechanism explaining job seeker attraction to
  organizations. Human Resource Management 56(4):629--649

\bibitem[{Fletcher et~al(1986)Fletcher, Danilovics, Fernandez, Peterson, and
  Reeder}]{fletcher1986attributional}
Fletcher GJ, Danilovics P, Fernandez G, Peterson D, Reeder GD (1986)
  Attributional complexity: An individual differences measure. Journal of
  Personality and Social Psychology 51(4):875

\bibitem[{Furia(2017)}]{furia2017}
Furia C (2017) What good is {B}ayesian data analysis for software engineering?
  In: Proceedings of the 39th IEEE/ACM International Conference on Software
  Engineering Companion, pp 374--376

\bibitem[{Furia et~al(2019)Furia, Feldt, and Torkar}]{furia2019bayesian}
Furia CA, Feldt R, Torkar R (2019) Bayesian data analysis in empirical software
  engineering research. IEEE Transactions on Software Engineering
  47(9):1786--1810

\bibitem[{Furia et~al(2021)Furia, Torkar, and Feldt}]{furia2021}
Furia CA, Torkar R, Feldt R (2021) Applying {B}ayesian analysis guidelines to
  empirical software engineering data: The case of programming languages and
  code quality. ACM Transactions on Software Engineering and Methodology
  forthcoming, accepted in October 2021.

\bibitem[{Furnham(1996)}]{furnham1996big}
Furnham A (1996) The {B}ig {F}ive versus the big four: the relationship between
  the {M}yers-{B}riggs type indicator ({MBTI}) and {NEO-PI} five factor model
  of personality. Personality and Individual Differences 21(2):303--307

\bibitem[{Furnham and Thorne(2013)}]{furnham2013stable}
Furnham A, Thorne JD (2013) Need for cognition. Journal of Individual
  Differences

\bibitem[{Gallagher(2012)}]{gallagher2012managing}
Gallagher VC (2012) Managing resources and need for cognition: Impact on
  depressed mood at work. Personality and Individual Differences 53(4):534--537

\bibitem[{{General Assembly of the World Medical
  Association}(2014)}]{general2014Helsinki}
{General Assembly of the World Medical Association} (2014) World medical
  association declaration of helsinki: ethical principles for medical research
  involving human subjects. The Journal of the American College of Dentists
  81(3):14

\bibitem[{Glass(2002)}]{glass2002facts}
Glass RL (2002) Facts and Fallacies of Software Engineering. Addison-Wesley
  Professional

\bibitem[{Gravetter and Forzano(2018)}]{gravetter2018research}
Gravetter FJ, Forzano LAB (2018) Research methods for the behavioral sciences.
  Cengage Learning

\bibitem[{Graziotin et~al(2014)Graziotin, Wang, and
  Abrahamsson}]{graziotin2014happy}
Graziotin D, Wang X, Abrahamsson P (2014) Happy software developers solve
  problems better: psychological measurements in empirical software
  engineering. PeerJ 2:e289

\bibitem[{Gren(2018)}]{gren2018standards}
Gren L (2018) Standards of validity and the validity of standards in behavioral
  software engineering research: the perspective of psychological test theory.
  In: Proceedings of the 12th ACM/IEEE International Symposium on Empirical
  Software Engineering and Measurement, ACM, p~55

\bibitem[{Gustavsson et~al(1997)Gustavsson, Weinryb, G{\"o}ransson, Pedersen,
  and {\AA}sberg}]{gustavsson1997stability}
Gustavsson JP, Weinryb RM, G{\"o}ransson S, Pedersen NL, {\AA}sberg M (1997)
  Stability and predictive ability of personality traits across 9 years.
  Personality and individual differences 22(6):783--791

\bibitem[{Haugtvedt and Petty(1992)}]{haugtvedt1992personality}
Haugtvedt CP, Petty RE (1992) Personality and persuasion: Need for cognition
  moderates the persistence and resistance of attitude changes. Journal of
  Personality and Social Psychology 63(2):308

\bibitem[{Heppner et~al(1983)Heppner, Reeder, and
  Larson}]{heppner1983cognitive}
Heppner PP, Reeder BL, Larson LM (1983) Cognitive variables associated with
  personal problem-solving appraisal: Implications for counseling. Journal of
  Counseling Psychology 30(4):537

\bibitem[{Hill et~al(2013)Hill, Foster, Elliott, Shelton, McCain, and
  Gouvier}]{hill2013need}
Hill BD, Foster JD, Elliott EM, Shelton JT, McCain J, Gouvier WD (2013) Need
  for cognition is related to higher general intelligence, fluid intelligence,
  and crystallized intelligence, but not working memory. Journal of Research in
  Personality 47(1):22--25

\bibitem[{Lins~de Holanda~Coelho et~al(2018)Lins~de Holanda~Coelho, Hanel, and
  Wolf}]{lins2018very}
Lins~de Holanda~Coelho G, Hanel PHP, Wolf LJ (2018) The very efficient
  assessment of need for cognition: Developing a six-item version. Assessment
  pp 1--16

\bibitem[{Hosio et~al(2020)Hosio, van Berkel, Oppenlaender, and
  Goncalves}]{Hosio2020CrowdsourcingDiets}
Hosio S, van Berkel N, Oppenlaender J, Goncalves J (2020) Crowdsourcing
  personalized weight loss diets. IEEE Computer 53(1):63--71

\bibitem[{{JASP Team}(2018)}]{jasp2018jasp}
{JASP Team} (2018) Jasp (version 0.11.1)[computer software]

\bibitem[{Jonason and Webster(2010)}]{jonason2010dirty}
Jonason PK, Webster GD (2010) The dirty dozen: A concise measure of the dark
  triad. Psychological Assessment 22(2):420

\bibitem[{Judge et~al(2009)Judge, Piccolo, and Kosalka}]{judge2009bright}
Judge TA, Piccolo RF, Kosalka T (2009) The bright and dark sides of leader
  traits: A review and theoretical extension of the leader trait paradigm. The
  Leadership Quarterly 20(6):855--875

\bibitem[{Jung(1931)}]{jung1931basic}
Jung CG (1931) Basic postulates of analytical psychology. Routledge

\bibitem[{Kearney et~al(2009)Kearney, Gebert, and Voelpel}]{kearney2009and}
Kearney E, Gebert D, Voelpel SC (2009) When and how diversity benefits teams:
  The importance of team members' need for cognition. Academy of Management
  Journal 52(3):581--598

\bibitem[{Kosinski et~al(2013)Kosinski, Stillwell, and
  Graepel}]{kosinski2013private}
Kosinski M, Stillwell D, Graepel T (2013) Private traits and attributes are
  predictable from digital records of human behavior. Proceedings of the
  National Academy of Sciences 110(15):5802--5805

\bibitem[{Kraichy and Chapman(2014)}]{kraichy2014tailoring}
Kraichy D, Chapman DS (2014) Tailoring web-based recruiting messages:
  Individual differences in the persuasiveness of affective and cognitive
  messages. Journal of Business and Psychology 29(2):253--268

\bibitem[{Lammers(1986)}]{lammers1986}
Lammers S (1986) Programmers at Work. Microsoft Press

\bibitem[{Leary and Hoyle(2009)}]{leary2009handbook}
Leary MR, Hoyle RH (2009) Handbook of individual differences in social
  behavior. Guilford Press

\bibitem[{Lee and Shneiderman(1978)}]{lee1978personality}
Lee JM, Shneiderman B (1978) Personality and programming: Time-sharing vs.
  batch preference. In: Proceedings of the Annual Conference on Software
  Engineering, IEEE, pp 561--569

\bibitem[{Lee and Vanpaemel(2018)}]{lee2018determining}
Lee MD, Vanpaemel W (2018) Determining informative priors for cognitive models.
  Psychonomic Bulletin \& Review 25(1):114--127

\bibitem[{Lee and Wagenmakers(2013)}]{lee2013}
Lee MD, Wagenmakers EJ (2013) Bayesian cognitive modeling: A practical course.
  Cambridge University Press

\bibitem[{Lee and Wagenmakers(2014)}]{lee2014bayesian}
Lee MD, Wagenmakers EJ (2014) Bayesian cognitive modeling: A practical course.
  Cambridge University Press

\bibitem[{Lenberg et~al(2015)Lenberg, Feldt, and
  Wallgren}]{lenberg2015behavioral}
Lenberg P, Feldt R, Wallgren LG (2015) Behavioral software engineering: A
  definition and systematic literature review. Journal of Systems and software
  107:15--37

\bibitem[{Li et~al(2020)Li, Ko, and Begel}]{li2020distinguishes}
Li PL, Ko AJ, Begel A (2020) What distinguishes great software engineers?
  Empirical Software Engineering 25(1):322--352

\bibitem[{Liang et~al(2008)Liang, Paulo, Molina, Clyde, and
  Berger}]{liang2008mixtures}
Liang F, Paulo R, Molina G, Clyde MA, Berger JO (2008) Mixtures of g priors for
  {B}ayesian variable selection. Journal of the American Statistical
  Association 103(481):410--423

\bibitem[{Machiavelli(2008)}]{machiavelli1532}
Machiavelli N (2008) The Prince. Hackett Publishing Company, Inc.

\bibitem[{Madrid and Patterson(2016)}]{madrid2016creativity}
Madrid HP, Patterson MG (2016) Creativity at work as a joint function between
  openness to experience, need for cognition and organizational fairness.
  Learning and Individual Differences 51:409--416

\bibitem[{Marreiros et~al(2017)Marreiros, Tonin, Vlassopoulos, and
  Schraefel}]{marreiros2017now}
Marreiros H, Tonin M, Vlassopoulos M, Schraefel M (2017) “now that you
  mention it”: A survey experiment on information, inattention and online
  privacy. Journal of Economic Behavior \& Organization 140:1--17

\bibitem[{Masegosa(2020)}]{masegosa2020learning}
Masegosa AR (2020) Learning under model misspecification: Applications to
  variational and ensemble methods. NeurIPS

\bibitem[{McCrae(1996)}]{mccrae1996social}
McCrae RR (1996) Social consequences of experiential openness. Psychological
  Bulletin 120(3):323--335

\bibitem[{McCrae and Costa~Jr(1989)}]{mccrae1989reinterpreting}
McCrae RR, Costa~Jr PT (1989) Reinterpreting the myers-briggs type indicator
  from the perspective of the five-factor model of personality. Journal of
  Personality 57(1):17--40

\bibitem[{McCrae and John(1992)}]{mccrae1992introduction}
McCrae RR, John OP (1992) An introduction to the five-factor model and its
  applications. Journal of Personality 60(2):175--215

\bibitem[{Mellblom et~al(2019)Mellblom, Arason, Gren, and
  Torkar}]{mellblom2019}
Mellblom E, Arason I, Gren L, Torkar R (2019) The connection between burnout
  and personality types in software developers. {IEEE} Software 36(5):57--64

\bibitem[{Miles(2014)}]{Miles2014VIF}
Miles J (2014) Tolerance and Variance Inflation Factor. American Cancer Society

\bibitem[{Myers(1976)}]{myers1976introduction}
Myers IB (1976) Introduction to type: A Guide to Understanding Your Results on
  the Myers-Briggs Type Indicator. CPP, Inc.

\bibitem[{van Nimwegen et~al(2006)van Nimwegen, Burgos, van Oostendorp, and
  Schijf}]{van2006paradox}
van Nimwegen CC, Burgos DD, van Oostendorp HH, Schijf HH (2006) The paradox of
  the assisted user: guidance can be counterproductive. In: Proceedings of the
  SIGCHI Conference on Human Factors in Computing Systems, pp 917--926

\bibitem[{Nuzzo(2014)}]{nuzzo2014scientific}
Nuzzo R (2014) Scientific method: statistical errors. Nature News 506(7487):150

\bibitem[{O'Brien(2007)}]{Obrien2007VIF}
O'Brien RM (2007) A caution regarding rules of thumb for variance inflation
  factors. Quality {\&} Quantity 41(5):673--690

\bibitem[{Palan and Schitter(2018)}]{palan2018prolific}
Palan S, Schitter C (2018) Prolific. ac—a subject pool for online
  experiments. Journal of Behavioral and Experimental Finance 17:22--27

\bibitem[{Park et~al(2008)Park, Baker, and Lee}]{park2008need}
Park HS, Baker C, Lee DW (2008) Need for cognition, task complexity, and job
  satisfaction. Journal of Management in Engineering 24(2):111--117

\bibitem[{Paulhus(1991)}]{paulhus1991}
Paulhus DL (1991) Measurement and control of response bias. In: Measures of
  Personality and Social Psychological Attitudes, Academic Press

\bibitem[{Paulhus and Williams(2002)}]{paulhus2002dark}
Paulhus DL, Williams KM (2002) The dark triad of personality: Narcissism,
  machiavellianism, and psychopathy. Journal of Research in Personality
  36(6):556--563

\bibitem[{Peer et~al(2017)Peer, Brandimarte, Samat, and
  Acquisti}]{peer2017beyond}
Peer E, Brandimarte L, Samat S, Acquisti A (2017) Beyond the turk: Alternative
  platforms for crowdsourcing behavioral research. Journal of Experimental
  Social Psychology 70:153--163

\bibitem[{Perry et~al(1994)Perry, Staudenmayer, and Votta}]{perry1994people}
Perry DE, Staudenmayer NA, Votta LG (1994) People, organizations, and process
  improvement. IEEE Software 11(4):36--45

\bibitem[{Petre and Blackwell(1997)}]{petre1997}
Petre M, Blackwell AF (1997) A glimpse of expert programmers' mental imagery.
  In: Proceedings of the Workshop on Empirical Studies of Programmers

\bibitem[{Petty and Jarvis(1996)}]{petty1996individual}
Petty RE, Jarvis WBG (1996) An individual differences perspective on assessing
  cognitive processes. Jossey-Bass

\bibitem[{Pohling et~al(2019)Pohling, Diessner, Stacy, Woodward, and
  Strobel}]{pohling2019moral}
Pohling R, Diessner R, Stacy S, Woodward D, Strobel A (2019) Moral elevation
  and economic games: The moderating role of personality. Frontiers in
  Psychology 10:1381

\bibitem[{Popoviciu et~al(2011)Popoviciu, Barbu, Costea, Culda, and
  Culda}]{popoviciu2011students}
Popoviciu SA, Barbu A, Costea D, Culda L, Culda S (2011) Student's desire to
  engage in cognitive activities, family of origin characterisitcs and need for
  cognition scores. Problems of Education in the 21st Century 33

\bibitem[{Ralph et~al(2021)Ralph, bin Ali, Baltes, Bianculli, Diaz, Dittrich,
  Ernst, Felderer, Feldt, Filieri, de~França, Furia, Gay, Gold, Graziotin, He,
  Hoda, Juristo, Kitchenham, Lenarduzzi, Martínez, Melegati, Mendez, Menzies,
  Molleri, Pfahl, Robbes, Russo, Saarimäki, Sarro, Taibi, Siegmund, Spinellis,
  Staron, Stol, Storey, Taibi, Tamburri, Torchiano, Treude, Turhan, Wang, and
  Vegas}]{ACMStd}
Ralph P, bin Ali N, Baltes S, Bianculli D, Diaz J, Dittrich Y, Ernst N,
  Felderer M, Feldt R, Filieri A, de~França BBN, Furia CA, Gay G, Gold N,
  Graziotin D, He P, Hoda R, Juristo N, Kitchenham B, Lenarduzzi V, Martínez
  J, Melegati J, Mendez D, Menzies T, Molleri J, Pfahl D, Robbes R, Russo D,
  Saarimäki N, Sarro F, Taibi D, Siegmund J, Spinellis D, Staron M, Stol K,
  Storey MA, Taibi D, Tamburri D, Torchiano M, Treude C, Turhan B, Wang X,
  Vegas S (2021) Empirical standards for software engineering research.
  \eprint{2010.03525}

\bibitem[{Rauthmann and Kolar(2012)}]{rauthmann2012dark}
Rauthmann JF, Kolar GP (2012) How “dark” are the dark triad traits?
  examining the perceived darkness of narcissism, machiavellianism, and
  psychopathy. Personality and Individual Differences 53(7):884--889

\bibitem[{Roberts and Mroczek(2008)}]{roberts2008stable}
Roberts BW, Mroczek D (2008) Personality trait change in adulthood. Current
  Directions in Psychological Science 17(1):31--35

\bibitem[{Roberts et~al(2007)Roberts, Kuncel, Shiner, Caspi, and
  Goldberg}]{roberts2007power}
Roberts BW, Kuncel NR, Shiner R, Caspi A, Goldberg LR (2007) The power of
  personality: The comparative validity of personality traits, socioeconomic
  status, and cognitive ability for predicting important life outcomes.
  Perspectives on Psychological Science 2(4):313--345

\bibitem[{Rouder and Morey(2012)}]{rouder2012default}
Rouder JN, Morey RD (2012) Default {B}ayes factors for model selection in
  regression. Multivariate Behavioral Research 47(6):877--903

\bibitem[{Russo(2021)}]{russo2021agile_success_model}
Russo D (2021) The agile success model: A mixed methods study of a large-scale
  agile transformation. ACM Transactions on Software Engineering and
  Methodology 30(4)

\bibitem[{Russo and Stol(2020)}]{russo2020gender}
Russo D, Stol KJ (2020) Gender differences in personality traits of software
  engineers. IEEE Transactions on Software Engineering forthcoming

\bibitem[{Russo et~al(2021{\natexlab{a}})Russo, Hanel, Altnickel, and van
  Berkel}]{russo2021daily}
Russo D, Hanel PHP, Altnickel S, van Berkel N (2021{\natexlab{a}}) The daily
  life of software engineers during the covid-19 pandemic. In: International
  Conference on Software Engineering, IEEE, pp 624--636

\bibitem[{Russo et~al(2021{\natexlab{b}})Russo, Hanel, Altnickel, and van
  Berkel}]{russo2020predictors}
Russo D, Hanel PHP, Altnickel S, van Berkel N (2021{\natexlab{b}}) Predictors
  of well-being and productivity among software professionals during the
  {COVID-19} pandemic--a longitudinal study. Empirical Software Engineering
  26(62):1--64

\bibitem[{Russo et~al(2021{\natexlab{c}})Russo, Hanel, and van
  Berkel}]{russo2021understanding}
Russo D, Hanel PHP, van Berkel N (2021{\natexlab{c}}) Understanding developers
  well-being and productivity: A longitudinal analysis of the covid-19
  pandemic. arXiv preprint arXiv:211110349

\bibitem[{Sadowski and Cogburn(1997)}]{sadowski1997need}
Sadowski CJ, Cogburn HE (1997) Need for cognition in the big-five factor
  structure. The Journal of Psychology 131(3):307--312

\bibitem[{Salzberger(2013)}]{salzberger2013attempting}
Salzberger T (2013) Attempting measurement of psychological attributes.
  Frontiers in Psychology 4:75

\bibitem[{Saucier(2009)}]{saucier2009recurrent}
Saucier G (2009) Recurrent personality dimensions in inclusive lexical studies:
  Indications for a big six structure. Journal of Personality 77(5):1577--1614

\bibitem[{Sharma and Stol(2020)}]{sharma2020}
Sharma GG, Stol KJ (2020) Exploring onboarding success, organizational fit, and
  turnover intention of software professionals. The Journal of Systems and
  Software 159

\bibitem[{Shneiderman(1980)}]{shneiderman1980software}
Shneiderman B (1980) Software psychology. Winthrop, Cambridge, Mass 48:161--172

\bibitem[{Shoaib et~al(2009)Shoaib, Nadeem, and Akbar}]{shoaib2009empirical}
Shoaib L, Nadeem A, Akbar A (2009) An empirical evaluation of the influence of
  human personality on exploratory software testing. In: Proceedings of the
  13th International Multitopic Conference, IEEE, pp 1--6

\bibitem[{Simmonds et~al(2018)Simmonds, Woods, and Spence}]{simmonds2018show}
Simmonds G, Woods AT, Spence C (2018) ‘show me the goods’: Assessing the
  effectiveness of transparent packaging vs. product imagery on product
  evaluation. Food Quality and Preference 63:18--27

\bibitem[{Singer et~al(1998)Singer, Mitchell, and
  Turner}]{singer1998consideration}
Singer M, Mitchell S, Turner J (1998) Consideration of moral intensity in
  ethicality judgements: Its relationship with whistle-blowing and
  need-for-cognition. Journal of Business Ethics 17(5):527--541

\bibitem[{Smith et~al(2001)Smith, Kerr, Markus, and
  Stasson}]{smith2001individual}
Smith BN, Kerr NA, Markus MJ, Stasson MF (2001) Individual differences in
  social loafing: Need for cognition as a motivator in collective performance.
  Group Dynamics: Theory, Research, and Practice 5(2):150

\bibitem[{Smith(1989)}]{smith1989personality}
Smith D (1989) The personality of the systems analyst: an investigation.
  Computer Personnel 12(2):12--14

\bibitem[{Stol and Fitzgerald(2018)}]{stol2018abc}
Stol KJ, Fitzgerald B (2018) The {ABC} of software engineering research. ACM
  Transactions on Software Engineering and Methodology 27(3):11

\bibitem[{Storey et~al(2020)Storey, Ernst, Williams, and
  Kalliamvakou}]{storey2020}
Storey MA, Ernst NA, Williams C, Kalliamvakou E (2020) The who, what, how of
  software engineering research: a socio-technical framework. Empirical
  Software Engineering 25(5):4097--4129

\bibitem[{Strobel et~al(2018)Strobel, Fleischhauer, Luong, and
  Strobel}]{strobel2018predicting}
Strobel A, Fleischhauer M, Luong C, Strobel A (2018) Predicting everyday life
  behavior by direct and indirect measures of need for cognition. Journal of
  Individual Differences

\bibitem[{Tuten and Bosnjak(2001)}]{tuten2001understanding}
Tuten TL, Bosnjak M (2001) Understanding differences in web usage: The role of
  need for cognition and the five factor model of personality. Social Behavior
  and Personality: an International Journal 29(4):391--398

\bibitem[{Valentine and Godkin(2019)}]{valentine2019moral}
Valentine S, Godkin L (2019) Moral intensity, ethical decision making, and
  whistleblowing intention. Journal of Business Research 98:277--288

\bibitem[{Venkatraman and Price(1990)}]{venkatraman1990differentiating}
Venkatraman MP, Price LL (1990) Differentiating between cognitive and sensory
  innovativeness: Concepts, measurement, and implications. Journal of Business
  Research 20(4):293--315

\bibitem[{Wagenmakers et~al(2018)Wagenmakers, Marsman, Jamil, Ly, Verhagen,
  Love, Selker, Gronau, {\v{S}}m{\'\i}ra, Epskamp
  et~al}]{wagenmakers2018bayesian}
Wagenmakers EJ, Marsman M, Jamil T, Ly A, Verhagen J, Love J, Selker R, Gronau
  QF, {\v{S}}m{\'\i}ra M, Epskamp S, et~al (2018) Bayesian inference for
  psychology. part i: Theoretical advantages and practical ramifications.
  Psychonomic Bulletin \& Review 25(1):35--57

\bibitem[{Weinberg(1971)}]{weinberg1971psychology}
Weinberg GM (1971) The psychology of computer programming. Van Nostrand
  Reinhold

\bibitem[{Weiner and Greene(2017)}]{weiner2017handbook}
Weiner IB, Greene RL (2017) Handbook of personality assessment. John Wiley \&
  Sons

\bibitem[{Wu et~al(2014)Wu, Parker, and De~Jong}]{wu2014need}
Wu CH, Parker SK, De~Jong JP (2014) Need for cognition as an antecedent of
  individual innovation behavior. Journal of Management 40(6):1511--1534

\bibitem[{Yamane(1973)}]{yamane1973statistics}
Yamane T (1973) Statistics: An introductory analysis. Harper \& Row

\end{thebibliography}

\pagebreak

\appendix

\section{Bayesian correlation and Bayesian Multi-model Linear Regression in JASP}
\label{app:BLR}

In Bayesian multi-model linear regression, JASP considers that the dependent variable (i.e., need for cognition) is only affected by a subset of the independent variables (i.e., personality traits). In that case, JASP evaluates all possible linear regression models defined by any of the subsets of the independent variables. As previously mentioned in the paper, JASP considers 512 alternative linear regression models. We refer to \citep{van2021tutorial} for a detailed explanation of how JASP defines this model.

When building the multi-model Bayesian linear regression in JASP, we use default options, and we considered the following priors:

\begin{itemize}
    \item \textbf{Model Prior:} We considered a uniform distribution over the 512 alternative linear regression models. We decided to use a uniform model prior because we did not have any prior knowledge that justifies a model that includes a smaller or larger subset of the personality traits for predicting is most plausible. JASP includes other model priors that favor regression models with a sparser number of personality traits, but, as we said before, we did not have any reasons for considering this kind of prior knowledge.  
        
    \item \textbf{Parameter Prior:} We considered the following four priors present in JASP: 
    the \textit{JZS prior} \citep{liang2008mixtures,rouder2012default}, the \textit{hyper-g prior} \citep{liang2008mixtures}, the \textit{hyper-g-Laplace prior} which is the same as the hyper-g prior but uses a Laplace approximation, and the \textit{hyper-g-n prior} which uses a hyper-g/n prior \citep{liang2008mixtures}. A justification for these prior is given by \citep{van2021tutorial}.

\end{itemize}

In the case of the Bayesian correlation analysis, we used different priors by changing the width parameter of the Beta prior. We took four values: 0.5, 1.0, 1.5, 2.0. These four values capture the range of values used by JASP when making robustness analysis for this prior\footnote{The specific guideline can be found at this URL: \url{http://static.jasp-stats.org/Manuals/Bayesian_Guide_v0.12.2.pdf}}.

Finally, in the Supplementary Material, we provided the raw results given by JASP for all these analysis, as well as several screenshots as a guide to reproduce the results reported in this work using JASP. 

\end{document}